\documentclass[11pt,preprint]{aastex}
\usepackage{epsfig}
\newcommand{\msun}{\mbox{${\rm M}_\odot$}}
\newcommand{\lsun}{\mbox{${\rm L}_\odot$}}

\begin{document}
\title{The Fate of Dwarf Galaxies in Clusters and
the Origin of Intracluster Stars. II. Cosmological Simulations}

\author{
Hugo Martel,\altaffilmark{1,2}
Paramita Barai,\altaffilmark{3} and
William Brito\altaffilmark{2}}

\altaffiltext{1}{D\'epartement de physique, de g\'enie physique et d'optique,
Universit\'e Laval, Qu\'ebec, QC, Canada}

\altaffiltext{2}{Centre de Recherche en Astrophysique du Qu\'ebec}

\altaffiltext{3}{Osservatorio Astronomico di Trieste, Trieste, Italy}

\begin{abstract}

We combine a N-body simulation 
algorithm with a subgrid treatment of galaxy formation,
mergers, and tidal destruction, and an observed conditional luminosity
function $\Phi(L|M)$, to study 
the origin and evolution of galactic and extragalactic
light inside a cosmological volume of size $(100\,{\rm Mpc})^3$, in a 
concordance $\Lambda$CDM model. This algorithm simulates the growth of
large-scale structures and the formation of clusters, the evolution of the
galaxy population in clusters, the destruction of galaxies by mergers and 
tides, and the evolution of the intracluster light. We find that 
destruction of galaxies by mergers dominates over destruction by tides by 
about an order of magnitude at all redshifts. However, tidal destruction is
sufficient to produce intracluster light fractions $f_{\rm ICL}$ that are 
sufficiently high to match observations. Our simulation produces 18 massive
clusters ($M_{\rm cl}>10^{14}\msun$) with values of $f_{\rm ICL}$ ranging from
1\% to 58\% at $z=0$. There is a weak trend of $f_{\rm ICL}$ to increase with
cluster mass. The bulk of the intracluster light ($\sim60\%$)
is provided by intermediate
galaxies of total masses $10^{11}\msun-10^{12}\msun$ and
stellar 
masses $6\times10^8\msun-3\times10^{10}\msun$ that were tidally destroyed
by even more massive galaxies. The contribution of low-mass galaxies to
the intracluster light is small and the contribution of dwarf galaxies
is negligible, even though, by numbers, most galaxies that are tidally 
destroyed are dwarfs.
Tracking clusters back in time, we find that their values of
$f_{\rm ICL}$ tend to increase over time, but can experience sudden changes
that are sometimes non-monotonic. These changes
occur during major mergers involving
clusters of comparable masses but very different intracluster luminosities.
Most of the tidal destruction events take place in the central regions of
clusters. As a result, the intracluster light is more centrally concentrated
than the galactic light.
Our results support tidal destruction of intermediate-mass galaxies as a
plausible scenario for the origin of the intracluster light.

\end{abstract}


\section{INTRODUCTION}

An important contribution to the total visible light emitted by massive,
X-ray galaxy clusters does not come from the galaxies themselves
\citep{zwicky51,arnaboldi04,lm04,feldmeieretal04a,feldmeieretal04b,
westetal95,
zibettietal05,gonzalezetal05,mihosetal05,kricketal06,kb07}.
This so-called {\it intracluster light} (ICL) is
attributed to intracluster stars, low surface brightness stars
located outside galaxies. Intracluster stars have been directly
observed \citep{fergusonetal98,arnaboldietal03,gal-yametal03,gerhardetal05}. 
These observations reveal that the intracluster stellar
population is diverse. It is mostly comprised of stars with masses of order
$1\msun$, ages up to $10^{10}{\rm yr}$, and metallicities $\rm[Fe/H]$ between
$-2$ and 0 \citep{williamsetal07}, but also includes red, old stars
\citep{kricketal06}
AGB stars \citep{durrelletal02}, planetary nebulae \citep{arnaboldietal96,
feldmeieretal03,arnaboldietal04,feldmeieretal04a}, novae and
supernovae.
\citep{tk77,ctw77,uson91,vilchez-gomez94,sk94,gal-yametal03,nso05}.
The fraction of stars that contribute to the ICL increases with the
mass of the clusters, and with the density of 
the environment: from loose groups ($<2\%$,
\citealt{castroetal03}), to Virgo-like ($10\%$,
\citealt{feldmeieretal04a,zibettietal05}) and rich clusters
($\sim20\%$ or higher, \citealt{tf95,feldmeieretal04b,kb07}). In the
cores of dense and rich clusters (like Coma) the local ICL fraction can be
as high as 50\% \citep{bernsteinetal95}.

Several models have been suggested to explain the origin of intracluster
stars. A comprehensive review of these various processes is provided by
\citet{tf11}. Essentially,
these models fall into four categories: intracluster star formation, ejection,
disruption of individual galaxies, or galactic interactions.

Some models for in-situ formation of intracluster stars have been suggested.
Gas-rich galaxies moving through the hot intracluster medium (ICM)
might experience 
ram-pressure stripping. The gas extracted from the galaxies will form
dense gaseous streams,
which under the right conditions can fragment to form stars
\citep{ft92,sunetal10}. This could explain the galactic
tails that are observed in some galaxies such as
ESO~137-001 and ESO~137-002. \citet{hatchetal08} observed a halo of diffuse
UV intergalactic light surrounding a radio galaxy (the Spiderweb Galaxy),
providing evidence for in-situ star formation outside galaxies.
Also, tidal stripping could provide another mechanism for extracting 
cold gas from galaxies. In the recent simulations of \citet{pssd10},
up to 30\% of intracluster stars formed in could gas clouds
stripped from substructures infalling into the cluster center.

Supernovae explosions inside close binary systems \citep{blaauw61} 
can produce high-velocity
stars, with velocities of several hundreds $\rm km/s$. 
Some of these stars 
could have enough kinetic energy to escape the gravitational potential 
of the parent galaxy and reach the intracluster space \citep{tf09}. 
However, as \citet{tf11} argue,
this is not an efficient mechanism for populating the intracluster space
with stars. Even
if ones assumes that every supernovae produces a high-velocity star, with
enough velocity to escape, the total number of intracluster stars produced
by this mechanism is simply too small. Three- or four-body encounters 
in dense stellar systems \citep{pra67} could also produce high-velocity 
runaway stars, and recent simulations suggest that these stars could reach
velocities as high as $400\,{\rm km\,s^{-1}}$ \citep{ggpz10}

A galaxy can get disrupted if it loses some of its
gravitational binding energy.
A gas-rich galaxy can become unbound by losing a significant fraction
of its gas. This could be caused by ram-pressure 
stripping, or by a galactic
wind powered by SNe explosions or an AGN. 
Another possible mechanism is the merger of
two galaxies which both host a central supermassive black hole. This will
likely lead to the formation of a single galaxy with a 
central binary black hole. If the binding energy of this binary black hole
exceeds the binding energy of the galaxy, the gravitational energy
extracted from the black holes will disrupt the galaxy \citep{tf11}.

While these various processes might contribute to some of the observed
ICL, it is generally accepted that most intracluster stars were formed
inside galaxies, and were later dispersed into the intracluster space by galaxy
interactions taking place during the evolution of the clusters.
This likely results from tidal stripping or tidal
destruction of galaxies during close
encounters (\citealt{weiletal97,gw98,gnedin03,willmanetal04,feldmeieretal04a,
rmmb06,cwk07,
pbz07}, \citealt{paperI}, hereafter Paper I, \citealt{ymvdb09,
wj09,rudicketal09,pssd10}), 
though an important contribution could also be provided
by stars ejected during galactic mergers \citep{muranteetal07}.
The ICL tends to be more concentrated than the galactic light 
\citep{aguerrietal05}, which is interpreted as evidence for the role
of galaxy collisions in the origin of the ICL \citep{zibettietal05}.
Notice that the higher rate of galaxy collisions
and higher ICM pressure found in the central regions of the clusters
would tend to increase the efficiency of most of the processes discussed
above, the exceptions being ejection of high-velocity stars (unaffected)
and SNe-driven galactic winds (possibly inhibited).

Several analytical and numerical studies of the origin and evolution of
the ICL have been performed. These include studies based on analytical 
modeling of galaxy formation and disruption \citep{pbz07}, N-body
simulations of large-scale structure formation combined
with an analytical prescription for galaxy formation, dispersion, and merging
\citep{napolitanoetal03,rmmb06,hbt08,rmmb11}, and hydrodynamical simulations 
\citep{willmanetal04,slrp05,muranteetal04,muranteetal07,pssd10,dmb10}.
In these numerical studies, there is always a trade-of 
between having good resolution or good statistics.
\citet{napolitanoetal03,willmanetal04,slrp05}, and \citet{rmmb06}
simulate either one cluster or a few clusters, so even
though these clusters are simulated with high resolution, they
might not be representative of the whole cluster
population. At the other extreme, \citet{muranteetal04,muranteetal07}
simulate a very large cosmological volume, containing
a statistically significant sample of clusters, but
cannot resolve the scale of
dwarf galaxies. In this work, we use an algorithm which
combines large-scale cosmological simulations with a 
semi-analytical treatment of mergers and tidal disruption.
This enables us to achieve
good statistics while resolving the processes responsible
for destroying dwarf galaxies. 

In this paper, we focus on the relative importance of the tidal destruction
and merger processes and their role in the evolution of the cluster
luminosities, and do not consider the properties of the ICM. In this case,
a full hydrodynamical simulation is not required, and we chose instead to
combine a N-body simulation with a subgrid treatment
of processes at galactic scales.
We use a high-resolution N-body simulation of large-scale structure formation,
as in \citet{hbt08} and \citet{rmmb11},
but with a 
different and complementary approach for galaxy formation, mergers,
and tidal destruction, as described in Paper~I (see \S~2.1 below). 
Our goals are to determine (1)
the fraction of galaxies of various masses destroyed by tides and mergers
during the formation and evolution of the clusters, (2) the contribution
of tidal destruction to the ICL, and (3) the brightness profile of the ICL
resulting from tidal destruction.


\section{THE NUMERICAL METHOD}

\subsection{N-body Simulation}

Simulating the formation and destruction of dwarf galaxies in a cosmological
context is quite challenging, because of the large dynamical range
involved. To get statistically significant results, we need to simulate a
volume of the universe large enough to contain several rich clusters. To
estimate this volume, we use the cluster mass function of \citet{bc93},
\begin{equation}
n_c(>\!M)\simeq4\times10^{-5}\left({M\over M^*}\right)^{-1}
e^{-M/M_*}h^3{\rm Mpc}^{-3}\,,
\end{equation}

\noindent where $M^*\simeq1.8\times10^{14}h^{-1}\msun$. Using $h=0.704$
and $M=10^{14}\msun$, we get $n_c=2.41\times10^{-5}{\rm Mpc}^{-3}$.
For a cubic volume of size $100\,{\rm Mpc}$, this gives 24 clusters
more massive than $M=10^{14}\msun$, which is probably sufficient to
get good statistics. 
In a $\Lambda$CDM universe with $\Omega_0=0.268$,
a $(100\,{\rm Mpc})^3$ volume contains a mass of 
$M_{\rm tot}=3.69\times10^{16}\msun$.
If we take the minimum mass of a dwarf galaxy to be $M_{\rm dw}=10^9\msun$,
we get $M_{\rm tot}/M_{\rm dw}=3.69\times10^7$. Lets assume that we perform
an N-body simulation with equal-mass particles, and that it takes
a minimum 100 particles per galaxy to properly resolve processes such as
galaxy merger and tidal destruction, we would then need 3.69~billion
particles. This is comparable to some of the largest N-body simulations
ever performed to this date, and would require an enormous investment
in human and computer resources.

The hydrodynamical simulations of \citet{muranteetal07} use three
different kinds of particles (dark matter, gas, and stars) with
masses $6.57\times10^9\msun$, $9.86\times10^8\msun$, $4.95\times10^8\msun$,
respectively. The gravity-only simulations of \citet{rmmb11} use particles
of mass $5\times10^8\msun$, while our own simulation uses particles
of mass $2.75\times10^8\msun$. \citet{hbt08} used the results of the
{\sl Millenium Simulation\/} \citep{springeletal05}, with particles
of mass $1.18\times10^9\msun$.
None of these simulations can resolve
the internal structure of dwarf galaxies and properly simulate the destruction
of these galaxies by mergers and tides.
To solve this problem, \citet{muranteetal07} use a group finder to
determine if particles belong to galaxies or are located in the
intracluster space (SKID, see \citealt{stadel01}).
\citet{rmmb11} used instead a standard ``zoom-in'' technique.
They first performed a relatively low-resolution simulation. They selected
a subset of massive clusters at redshift $z=0$, and ran the simulation
a second time, with more resolution inside the regions where these clusters
formed. \citet{pssd10} used the same approach, by selecting a sample of 16
clusters from the {\sl Millenium Simulation\/} and resimulating them with
higher resolution.
This approach provides very-high resolution at reasonable computational
cost. However, only a few clusters are being
simulated at that high resolution.
\citet{hbt08} use a semi-analytical model to describe galaxy formation
and disruption.
In this paper, we use an approach that we first introduced
in Paper~I. We represent each galaxy in the system
(regardless of its mass) using {\it one single particle}. In this
approach, the merger and tidal destruction of galaxies cannot be
directly simulated, but instead are treated in the algorithm
as subgrid physics. When two particles representing galaxies come close
to each other, we can calculate the gravitational potential energy
between them. We can also calculate the tidal field caused by one
galaxy at the location of the other. With these, we can set rules
that dictate when mergers and tidal destruction take place.
This is fairly crude compared to an actual simulation of the mergers
and tidal destruction events, {\it but is expected 
to make statistically correct
predictions for a large number of events.\/}
The main advantage of this approach is that it does not rely on zoom-ins,
and thus enables us to simulate a larger number of clusters at high
resolution.
This approach was developed
and tested on isolated clusters (Paper~I). In this paper, we apply the 
same approach to a cosmological simulation, which enables us to follow
the formation and evolution of a statistically significant number of clusters.

We consider a concordance $\Lambda$CDM model with $\Omega_0=0.268$,
$\lambda_0=0.732$, and $h=0.704$.
We perform a high-resolution
simulation in a $(100\,{\rm Mpc})^3$ comoving box with
periodic boundary conditions, using a Particle-Mesh (PM) algorithm 
with $512^3$ particles and a $1024^3$ mesh. 
The total mass in the box is $M_{\rm tot}=3.686\times10^{16}\msun$
and the mass per particle is $M_{\rm part}=2.747\times10^{8}\msun$.
The length resolution is $97.7\,\rm kpc$ comoving. We assume that dwarf 
galaxies of mass $M_{\min}=2\times10^9\msun$
form inside cells where the density is 200 times the mean density.
Each galaxy is represented by a ``galaxy particle.'' These particles are
treated like PM particles, but have the ability to form, merge,
and get tidally destroyed. The treatment of these processes by the
algorithm is described in the following sections.

\subsection{Formation of Dwarf Galaxies}

To include galaxy formation in our N-body simulations, we assume
that dwarf galaxies of mass $M=M_{\min}$ form by monolithic collapse,
while more massive galaxies form by the merger of smaller galaxies.
Therefore, in our code implementation, the formation of massive
galaxies by mergers is handled by the merging module,
and the galaxy formation module only handles the formation of galaxies
of mass $M=M_{\min}$.

The computational volume is divided into $1024^3$ PM cells.
We assume that dwarf galaxies form in cells where the density 
$\rho_X^{\phantom1}$ of 
background matter exceeds a  threshold density 
$\rho_{\rm GF}^{\phantom1}=\Delta_c\bar\rho(z)$, 
where $\bar\rho(z)$ is the mean density
of the universe at redshift $z$, and $\Delta_c=200$.
We use the density of the background matter, and not
the total density, because the matter already locked up in galaxies
is unavailable to form new galaxies, except by mergers.
We assume that in 
each cell that satisfies this criterion,
there is a probability $P$ of forming
a dwarf galaxy of mass $M_{\min}$ during a time interval $\Delta t$,
given by $P=\Psi\Delta t$, where $\Psi$ is a galaxy formation rate.
The number of galaxies created during a timestep is therefore
\begin{equation}
N_{\rm gal}=\Psi N_{\rm cell}\Delta t\,.
\label{GF}
\end{equation}

\noindent where $N_{\rm cells}$ is the number of cells that satisfy the
criterion. We adjust the value of $\Psi$ by requiring that 
the galaxy luminosity function at $z=0$ is
consistent with observations (see \S3.1 below). 

We select a subset of $N_{\rm gal}$ cells randomly among the $N_{\rm cell}$ cells
that satisfy the criterion $\rho_X^{\phantom1}>\rho_{\rm GF}^{\phantom1}$, 
and we create a galaxy of mass $M_{\min}$ in each of these cells.
To do so, we consider a Gaussian density profile:
\begin{equation}
\rho(r)=\rho_{\rm GF}^{\phantom1}e^{-r^2/2w^2}\,,
\end{equation}

\noindent where $r$ is the distance from the center of the cell, and the
width
$w$ is defined by $(2\pi)^{3/2}\rho_{\rm GF}^{\phantom1}w^3=M_{\min}$. 
This profile contains
a total mass $M_{\min}$. We identify all dark matter particles located within
a distance $r=4w$ from the center of the cell. We then remove from each
particle a mass $\Delta m=Ce^{-r^2/2w^2}$, where the constant $C$ is adjusted
such that the total mass removed is equal to $M_{\min}$. Instead of locating
the newborn galaxy in the exact center of the cell, we calculate the 
center-of-mass position and velocity of the matter that was removed
from dark matter particles, and these become the position and velocity
of the galaxy, respectively. This ensures that mass and momentum
are conserved. The initial radius of the galaxy is set to
$s=(3M_{\min}/4\pi\rho_{\rm GF}^{\phantom1})^{1/3}$, the radius of a uniform
sphere with $M=M_{\min}$ and $\rho=\rho_{\rm GF}^{\phantom1}$.

For the simulation presented in this paper, we used
a minimum mass $M_{\min}=2\times10^9\msun$, which corresponds to the
mass of 7 dark matter particles.
The corresponding filter width is $w=25.8\,{\rm kpc}$, or 0.2645 PM cells.

\subsection{The Subgrid Physics}

Our treatment of subgrid physics is presented in Paper I, and we refer 
the reader to that paper for details. Here we briefly summarize the 
method used. At each timestep, we identify all pairs of galaxies that are
sufficiently close that the center of one galaxy is inside the
other galaxy, that is, $r_{ij}<\max(s_i,s_j)$, where $r_{ij}$ is the
separation between galaxies $i$ and $j$, and $s_i$, $s_j$ are their radii.
For each pair, we calculate the total energy of the two
galaxies in the center-of-mass frame,
\begin{equation}
E_{ij}=K_i+K_j+W_{ij}+U_i+U_j\,,
\label{energy}
\end{equation}

\noindent where $K_i$ and $K_j$ are the kinetic energy of galaxies $i$ and $j$
in the center-of mass frame, $U_i$ and $U_j$ are 
the internal energies of the galaxies, and
$W_{ij}=-Gm_im_j/r_{ij}$ is the gravitational
potential energy of the galaxy pair,
respectively. The internal energies
depend on the masses and radii of the galaxies. They are given by 
\begin{equation}
U_i=-{\zeta Gm_i^2\over2s_i}\,,
\end{equation}

\noindent where $m_i$ and $s_i$ are the mass and radius of galaxy $i$, 
respectively. The {\it geometric factor} $\zeta$ depends on the density
profile of the galaxy, but does not vary much for any reasonable profile,
so, as in Paper~I, we use $\zeta=1$, which is the correct value for
a truncated isothermal sphere in virial equilibrium.

If $E_{ij}<0$, the merger criterion is satisfied 
and the two galaxies merge, to form a new galaxy $k$. The mass, position,
velocity, and radius of that galaxy are initialized using:
\begin{eqnarray}
m_k&=&m_i+m_j\,,\\
{\bf r}_k&=&{\bf r}_{ij}\,,\\
{\bf v}_k&=&{\bf v}_{ij}\,,\\
s_k&=&-{\zeta Gm_k^2\over2E_{ij}}={\zeta Gm_k^2\over2|E_{ij}|}\,,
\end{eqnarray}

\noindent where ${\bf r}_{ij}$ and ${\bf v}_{ij}$ are the center-of-mass 
position and velocity, respectively. These equations ensure conservation
of mass, momentum, and energy during mergers.

Tidal disruption is more tricky. When a galaxy is disrupted by the tidal field
of a more massive one, the inner part of the galaxy might survive, while the
outer part gets stripped. Some of the material stripped might then escape the
system, or might get accreted by the massive galaxy. We simplify the
problem by using an all-or-nothing approach.
We identify pairs of galaxies which are sufficiently close that the
separation between their {\it edges\/} is smaller than the radius of the largest
galaxies, that is, $r_{ij}-s_i-s_j<\max(s_i,s_j)$.
Let us assume that galaxy $i$ is the most massive of the pair.
We compare the tidal field of galaxy $i$
at the location of galaxy $j$ with the gravitational field of galaxy $j$
itself. If we estimate that the tidal field is strong enough to unbind more
than 50\% of the mass of galaxy $j$, then the tidal disruption criterion is
satisfied, and we assume that galaxy $j$ is totally
destroyed. Otherwise, galaxy $j$ survives the encounter.
In the case of a tidal destruction, we also check the merger criterion
[eq.~(\ref{energy})].
If that criterion is also satisfied, we assume that galaxy $j$ is destroyed by
the tidal field of galaxy $i$, and then the fragments accrete onto galaxy $i$.
Numerically this is treated as a merger. Galaxy particles 
being destroyed and not reaccreted are flagged, to indicate that they
do not represent galaxies anymore but rather tidal fragments.
They remain in the simulation, but are ignored during
subsequent encounters. This enables us to track the motion of tidal
fragments, and eventually determine in which cluster they end up
(see \S~3.3 below).

As we argue in Paper~I, our treatment of mergers and tidal disruption would 
be too simplistic to describe individual events, but can be used to
describe the net, collective effect of tens of thousands of galactic
encounters.

\section{RESULTS}

\subsection{Galaxy Luminosity Function and Stellar Mass function}

\begin{figure}
\begin{center}
\includegraphics[width=6in]{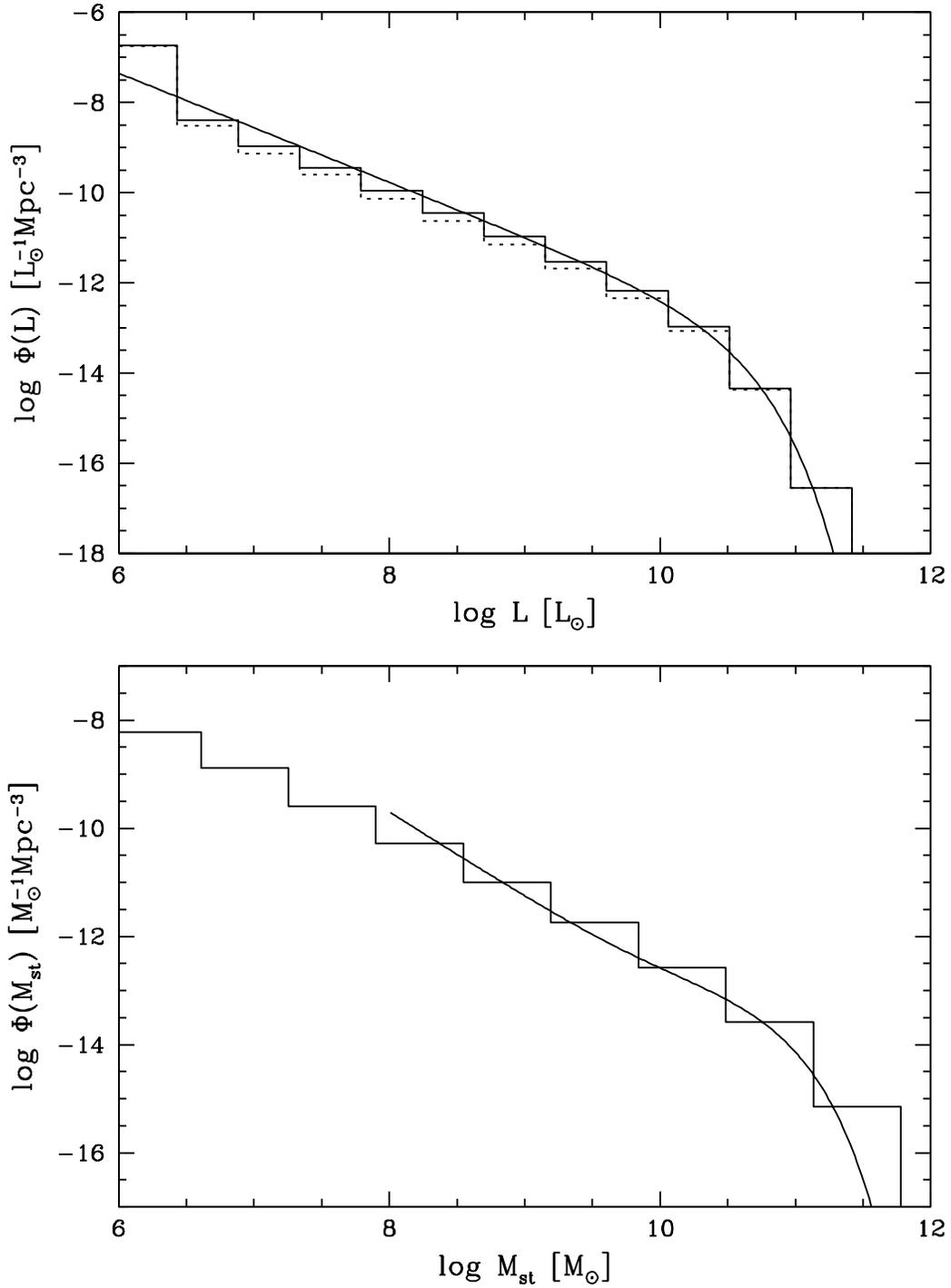}
\caption{Top panel: Luminosity function of galaxies at $z=0$.
Dotted histogram: central galaxies only; solid histogram: all galaxies;
solid curve: Schechter luminosity function with
$\phi^*=1.61\times10^{-2}h^3{\rm Mpc}^{-3}$,
$L^*=9.64\times10^9h^{-2}\lsun$ and $\alpha=-1.21$.
Bottom panel: Stellar mass function of galaxies at $z=0$.
Histogram: all galaxies; solid curve: Stellar mass
function of \citet{bgd08} (their eqs.~[2]-[3]).}
\label{histo2}
\end{center}
\end{figure}

The simulation produces 79,751 galaxies with masses in the range 
$2\times10^9\msun-7.47\times10^{13}\msun$,
including 251 objects with mass $M>10^{12}\msun$. Clearly, such massive
objects cannot
be individual galaxies. We interpret them as individual subhalos hosting
several galaxies, more specifically a {\it central galaxy\/} and
one or several {\it satellite galaxies}. With this in mind,
we can now calculate the luminosity function of galaxies.
To convert masses into luminosities, we use the conditional luminosity
function of \citet{ymvdb03}. The average number of galaxies in the
luminosity interval $[L-dL/2,L+dL/2]$ hosted by a halo of mass $M$
is given by
\begin{equation}
\phi(L|M)={\tilde\phi^*\over\tilde L^*}
\left({L\over\tilde L^*}\right)^{\tilde\alpha}e^{-L/\tilde L^*}\,,
\label{philm}
\end{equation}

\noindent where $\tilde\phi^*$, $\tilde L^*$, and $\tilde\alpha$ are
functions of the halo mass. The average 
mass-to-light ratio of halos is approximated by the following fitting formula:
\begin{equation}
{M\over L}={1\over2}\left({M\over L}\right)_0
\left[\left({M\over M_1}\right)^{-\beta}
+\left({M\over M_1}\right)^{\gamma^{\phantom0}_1}\right]\,.
\label{ml1}
\end{equation}

\noindent
If a halo hosts more than one galaxy,
the mean luminosity of the most luminous galaxy is given by
\begin{equation}
\bar L_c(M)=\tilde\phi^*\tilde L^*\Gamma(\tilde\alpha+2,L_1/\tilde L^*)\,,
\label{lc}
\end{equation}

\noindent where $\Gamma$ is the incomplete Gamma function, and $L_1$ is
defined by
\begin{equation}
\tilde\phi^*\Gamma(\tilde\alpha+1,L_1/\tilde L^*)=1\,.
\end{equation}

\noindent 
Note that there are more recent studies of the halo mass function
and stellar mass function, from which the mass-to-light ratio
can be calculated \citep{tinkeretal08,bcw10}. 
But these studies are limited to halos
of mass $M>10^{11}\msun$. The $M/L$ ratio of \citet{ymvdb03}
covers the range $M=10^9h^{-1}\msun-10^{14}h^{-1}\msun$,
which is appropriate for our work.
We used their model M1, defined by $M_1=10^{11.27}h^{-1}\msun$,
$(M/L)_0=134h\msun/\lsun$, $\beta=0.77$, $\gamma^{\phantom0}_1=0.32$.

Using equation~(\ref{lc}), we calculated the central luminosity $\bar L_c$
of each galaxy particle in the simulation.
The resulting luminosity function is shown by the dotted histogram in the
top panel of Figure~\ref{histo2}. 
For comparison, the solid curve shows a Schechter
luminosity function with $\phi^*=1.61\times10^{-2}h^3{\rm Mpc}^{-3}$,
$L^*=9.64\times10^9h^{-2}\lsun$ and $\alpha=-1.21$ \citep{ymvdb03}.
There is an excellent agreement at high luminosities, $L>10^{10}\lsun$.
At lower luminosities, our simulated luminosity function is systematically 
below the Schechter function, except in the lowest bin where our
results exceed the Schechter function by an order of magnitude.
However, if we interpret galaxy particles of mass $M>10^{12}\msun$ as
actually being halos containing several galaxies, then equation~(\ref{lc})
underestimates the total luminosity of these objects. By integrating
equation~(\ref{philm}), we can calculate the contribution of satellite
galaxies in each luminosity bin, and include it in the luminosity
function. The result is shown by the solid histogram in the top panel of
Figure~\ref{histo2}. There is now a good agreement with the Schechter
function at all luminosities $L>10^{8.5}\lsun$, while there is still a 
mismatch at lower luminosities. We attribute this to the discreteness of the
algorithm, in which all galaxy masses are multiples of $M_{\min}$.
In particular, the three lowest luminosity bins in Figure~\ref{histo2}
contain, respectively, galaxies of mass $M_{\min}$, of mass $2M_{\min}$, and
of masses $3M_{\min}$ and $4M_{\min}$. We can understand the large excess in the
lowest luminosity bin by noting that galaxies of mass $M_{\min}$ are allowed
to form directly, while more massive galaxies must be built-up through a
series of mergers.

We find this agreement quite remarkable. There is nothing in our
galaxy formation algorithm that guaranties a priori that the final
luminosity function would even resemble a Schechter function.
The model was never tuned to obtain this result, and there is not much that
{\it could\/} be tuned. 
We use a density threshold of $200\bar\rho$ for identifying
collapsed objects, which is a standard value based on theoretical 
arguments.\footnote{This is an approximation to $18\pi^2$, the exact
value for a spherical collapse in a $\Omega_0=1$ universe.}
The geometric factor $\zeta$ entering into the merger criterion
is a free parameter, but its value is close to unity for any reasonable
density profile. We argue that the features seen in the simulated luminosity 
function at low luminosities are caused by the discreteness of the galaxy 
masses. Hence, using a different value of $M_{\min}$ would most likely move
these features to different luminosities without affecting the high-end of the
luminosity function. The only real free parameter in our model is the
galaxy formation rate $\Psi$. Changing that parameter might improve the
fit at low luminosities, at the cost of worsening it at high luminosities.
Overall, we find that our model provides a satisfactory fit to the Schechter
luminosity function. In particular, it correctly predicts the value of the
luminosity break $L_*$, though that value (or the corresponding mass $M_*$)
is not built-in in the model.

Since the $M$ vs $L$ relation is non-linear, when two galaxies merge, the
merger remnant does not have the total luminosity of the two progenitors.
This simply reflects the fact that galactic evolution is much more than
just a series of mergers. The luminosity of galaxies is affected by processes
such as star formation and evolution, accretion, galactic winds, AGN activity,
and so on. Since these processes are not included in our model, we cannot
predict the evolution of the luminosities of galaxies during and between 
mergers. This is why we calculate
the values of the luminosity using the observed
$M$ vs. $L$ relation.

Once we have the luminosities,
we can easily estimate the stellar masses. 
We use the relation given by \citet{belletal03} for g-band
luminosities: $\log_{10}(M_{\rm st}/L)=-2.61+0.998\log_{10}M_{\rm st}h^2$, where
the stellar mass $M_{\rm st}$ and luminosity $L$ are in solar units.
For a cosmological model with $h=0.704$, this reduces to 
$M_{\rm st}=0.000142L^{1.425}$.
The resulting galactic stellar mass function is shown by the 
histogram in the bottom panel of Figure~\ref{histo2}. For comparison,
we show the observed stellar mass function of \citet{bgd08}, which covers
the stellar mass range $M_{\rm st}=10^8-10^{12}\msun$. There is an
excellent agreement over the range of masses covered by the observations,
except possibly at the highest mass bin. 

\subsection{Global Properties}

\begin{figure}
\begin{center}
\includegraphics[width=6.5in]{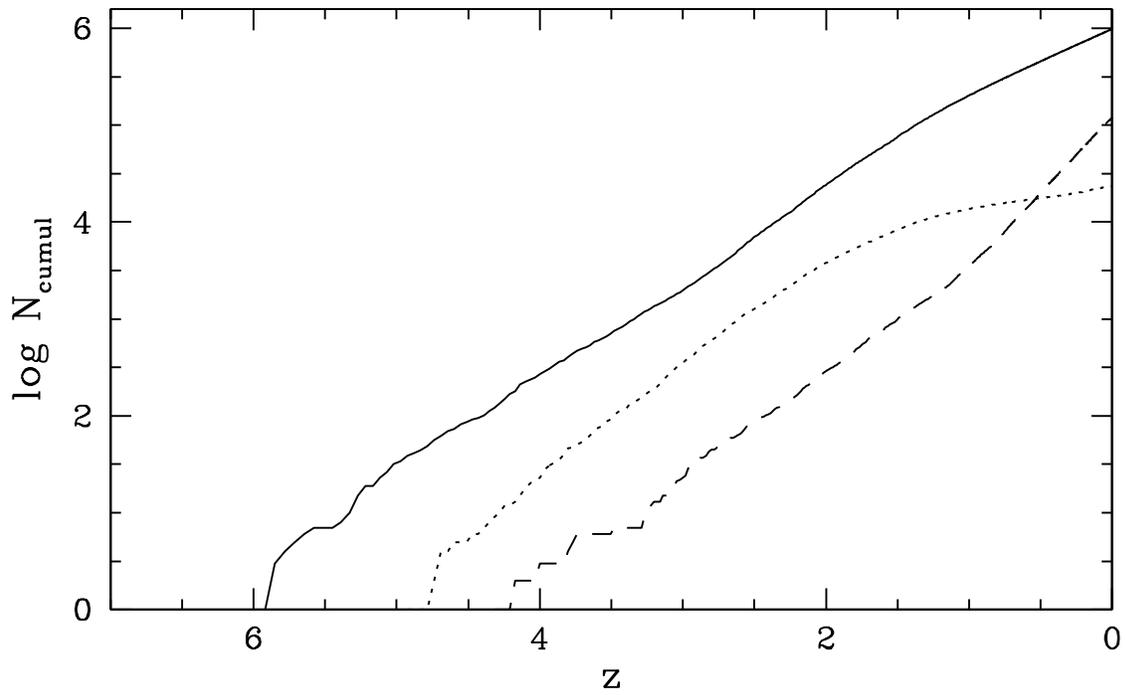}
\caption{
Cumulative number of mergers (solid line), tidal destruction events,
with fragments dispersed in the intracluster space (dotted line), and tidal
destruction events followed by accretion of the fragments onto the
massive galaxy (dashed line), versus redshift.}
\label{destruct}
\end{center}
\end{figure}

Figure~\ref{destruct} shows the cumulative number of mergers,
of tidal destruction events with the tidal fragments dispersed in the
intracluster space, 
and of tidal destruction events followed by accretion of the fragments 
onto the massive galaxy (these cases are also counted as mergers).
The number of all types of events increase roughly exponentially with redshift.
The delay between the start of merger events ($z=5.9$)
and tidal destruction events ($z=4.8$)
reflects the time it takes to build galaxies of different masses,
an essential condition for tidal destruction.
At redshifts $z>1$, more than 90\% of tidal destruction events result in
the fragments being dispersed into the intracluster space, 
and therefore contributing
to the ICL. After $z=1$, accretion of tidal fragments by the massive
galaxy becomes more common, and dominates after $z=0.5$.

\begin{figure}
\begin{center}
\includegraphics[width=6.5in]{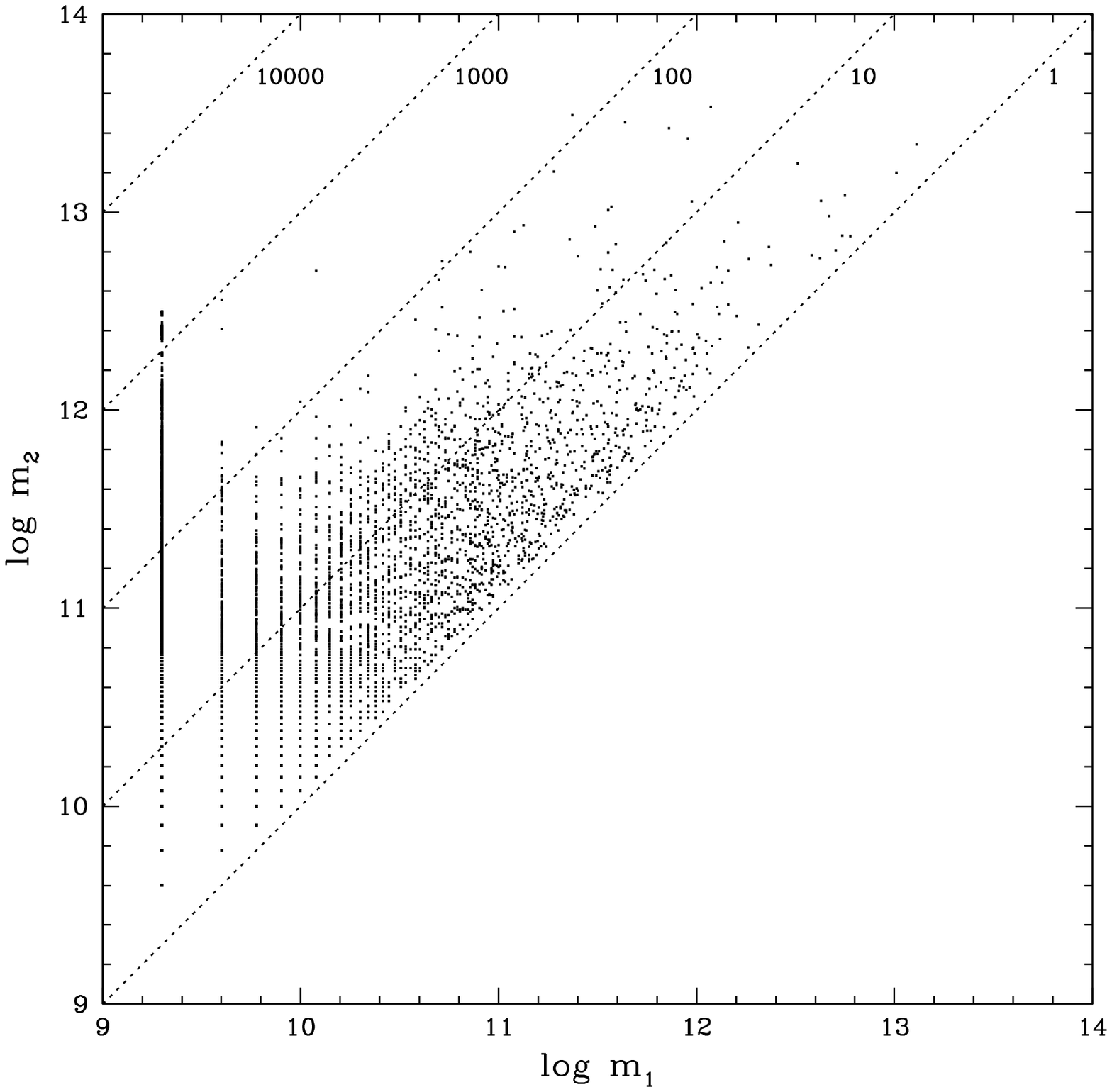}
\caption{
Encounters resulting in tidal destruction,
with dispersion of the fragments. $m_1$ and $m_2$ are the mass
of the lower- and higher-mass galaxies, respectively. Dashed lines indicate
mass ratios.}
\label{tides_mass}
\end{center}
\end{figure}

\begin{figure}
\begin{center}
\includegraphics[width=6.5in]{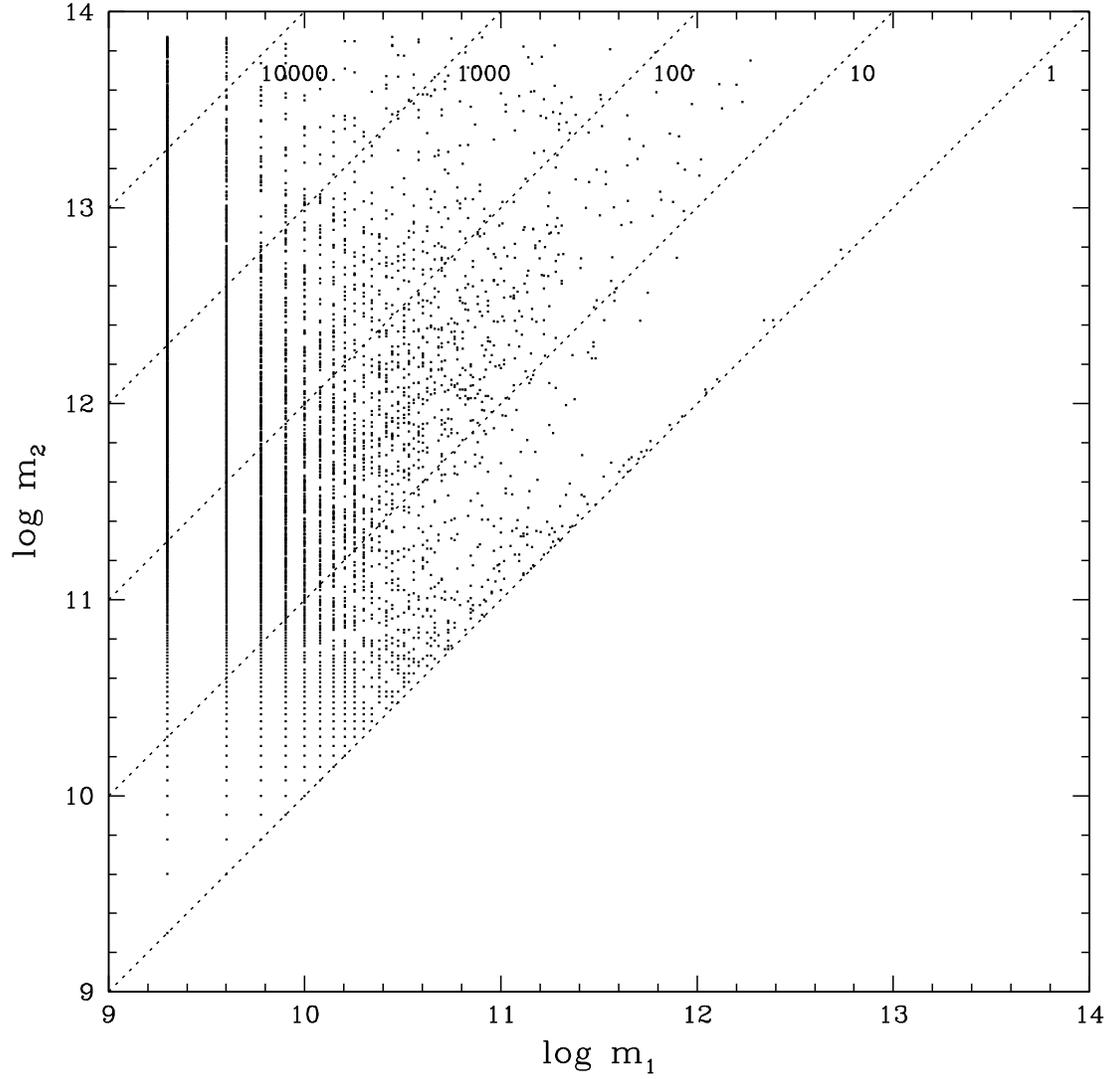}
\caption{
Encounters resulting in mergers (including tidal destruction
followed by accretion). $m_1$ and $m_2$ are the mass
of the lower- and higher-mass galaxies, respectively. Dashed lines indicate
mass ratios.}
\label{merge_mass}
\end{center}
\end{figure}

\begin{figure}
\begin{center}
\includegraphics[width=6.5in]{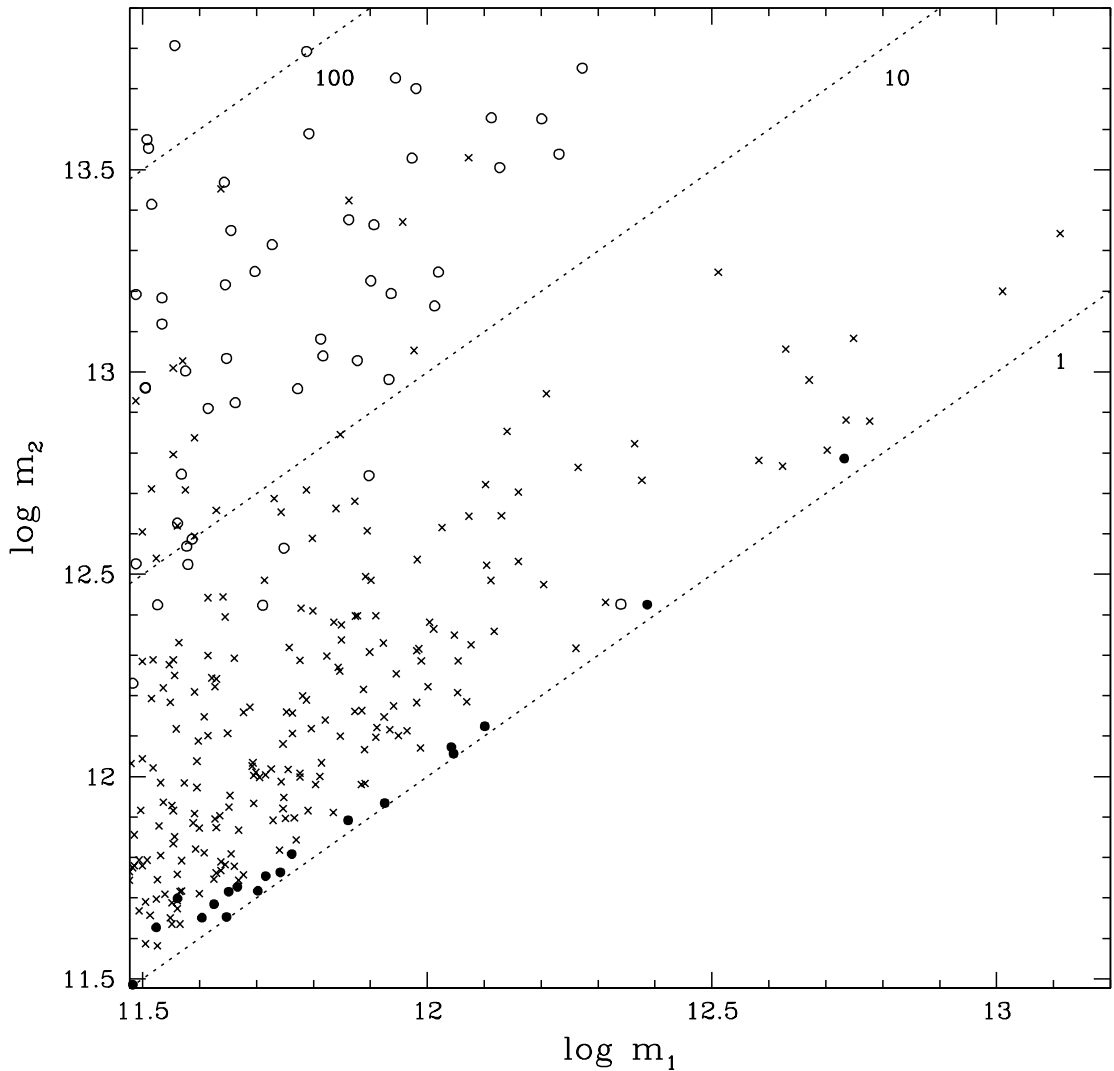}
\caption{
Encounters involving galaxies with masses $m>3\times10^{11}\msun$. 
$m_1$ and $m_2$ are the mass
of the lower- and higher-mass galaxies, respectively. 
Solid circles: direct mergers; crosses: tidal destruction
with fragments dispersed into the intracluster space; open circles: tidal
destruction followed by accretion.
Dashed lines indicate mass ratios.}
\label{encounters_himass}
\end{center}
\end{figure}

Figure~\ref{tides_mass} shows the mass of the galaxies involved in tidal
destruction followed by
dispersion, with $m_1$ being the mass of the galaxy being destroyed, and
$m_2$ being the mass of the other, more massive galaxy. The most striking 
feature is that tidal destruction is not limited to dwarf galaxies, but
covers more than three orders of magnitudes in mass.
The most massive galaxy destroyed had a total
mass $m_1=1.29\times10^{13}\msun$,
and was destroyed by a galaxy of mass $m_2=2.20\times10^{13}\msun$ 
at redshift $z=0.48$. 
Figure~\ref{merge_mass} shows a similar plot, for all merger events
(direct mergers, and tidal destruction followed by accretion). There are
noticeable differences. First, the top left corner of the plot is
populated, showing that encounters with very large mass ratios
($m_2/m_1>2500$) never result in dispersion, but can result in mergers.
Second, there is a clear ``desert'' in Figure~\ref{merge_mass}
at masses $m_1>3\times10^{11}\msun$ (or $M_{\rm st}>5\times10^9\msun$),
between mass ratios $m_2/m_1=1$ and 10, which is not found in
Figure~\ref{tides_mass}. To investigate this, we focus on encounters
with $m_1>3\times10^{11}\msun$, and combine the two figures. The results are
shown in Figure~\ref{encounters_himass}, where we use different symbols
for direct mergers, tidal destruction followed by dispersion,
and tidal destruction followed by accretion. There is a remarkable separation
between the three processes. Direct mergers occur at low mass ratios,
$m_2/m_1<1.2$, dispersion occurs mostly at intermediate ratios,
$1.2<m_2/m_1<10$, and accretion occurs mostly at high mass ratio, $m_2/m_1>10$.
To explain these results, let us consider what happens, in the 
simulation, during an encounter between a
galaxy of mass $m_1$ and size $s_1$ and a galaxy of mass $m_2>m_1$ and
size $s_2$, separated by a distance $r_{12}$. 
The tidal disruption and merger criteria used
in our model will determine the outcome of this encounter.
The gravitational field that
binds the first galaxy is of order $E\sim Gm_1/s_1^2$, while the tidal field
of the second galaxy at the location of the first one is of order
$T\sim Gm_2s_1/r_{12}^3$. The ratio of these fields, $T/E$ 
is then proportional to
$(m_2/m_1)(s_1/r_{ij})^3$. Since $s_1<r_{ij}$, it takes a certain mass ratio
for tidal destruction to occur. Hence, encounters between galaxies of
comparable masses can only result in either mergers (solid circles in
Fig.~\ref{encounters_himass}) or nothing. But if the mass ratio
is sufficiently large for the tidal disruption criterion to be
satisfied, the lower-mass galaxy will be destroyed, 
and the merger criterion will
determine whether the fragments are dispersed, or accreted by the
higher-mass galaxy. Unlike the tidal disruption criterion, the merger
criterion depends on
the velocity of the galaxies [$K$-terms in eq.~(\ref{energy})]. High-mass
galaxies tend to be located in massive clusters. Their velocities
are usually not determined by their properties and the
ones of their immediate neighbors, but rather by the overall properties
of the cluster in which these galaxies are located. In particular, we expect
the velocity of a galaxy orbiting inside a cluster to be of order the
velocity dispersion at its location, independently of its mass of the mass
of its neighbors.  We can then rewrite equation~(\ref{energy}) as
\begin{equation}
E_{12}\approx{m_1\sigma^2\over2}+{m_2\sigma^2\over2}-{Gm_1m_2\over r_{12}^2}
-{\zeta Gm_1^2\over s_1}-{\zeta Gm_2^2\over s_2}\,,
\label{energy2}
\end{equation}

\noindent where $\sigma$ is the local velocity dispersion. If $m_1$ and $m_2$
are small, the kinetic energy terms will dominate, and the criterion will
fail ($E_{12}>0$). The galaxies will simply pass by each other without merging,
and if the tidal disruption
criterion is satisfied (notice that it does not depend on
velocities), the lower-mass galaxy will be destroyed and the fragments
will be dispersed (crosses in Fig.~\ref{encounters_himass}). 
If we then keep $m_1$ constant and increase $m_2$, the second, third, and
fifth terms in equation~(\ref{energy2}) will increase in amplitude.
With two of these terms being negative, and the last one being quadratic
in $m_2$, $E_{12}$ will decrease, and for high enough values of $m_2$,
the criterion $E_{12}<0$ will be satisfied. The galaxies will merge,
and since increasing $m_2$ while keeping $m_1$ constant also favors the
tidal disruption criterion, the outcome will be tidal destruction followed by
reaccretion of the fragments (open circles in Fig.~\ref{encounters_himass}).

\begin{figure}
\begin{center}
\includegraphics[width=6.5in]{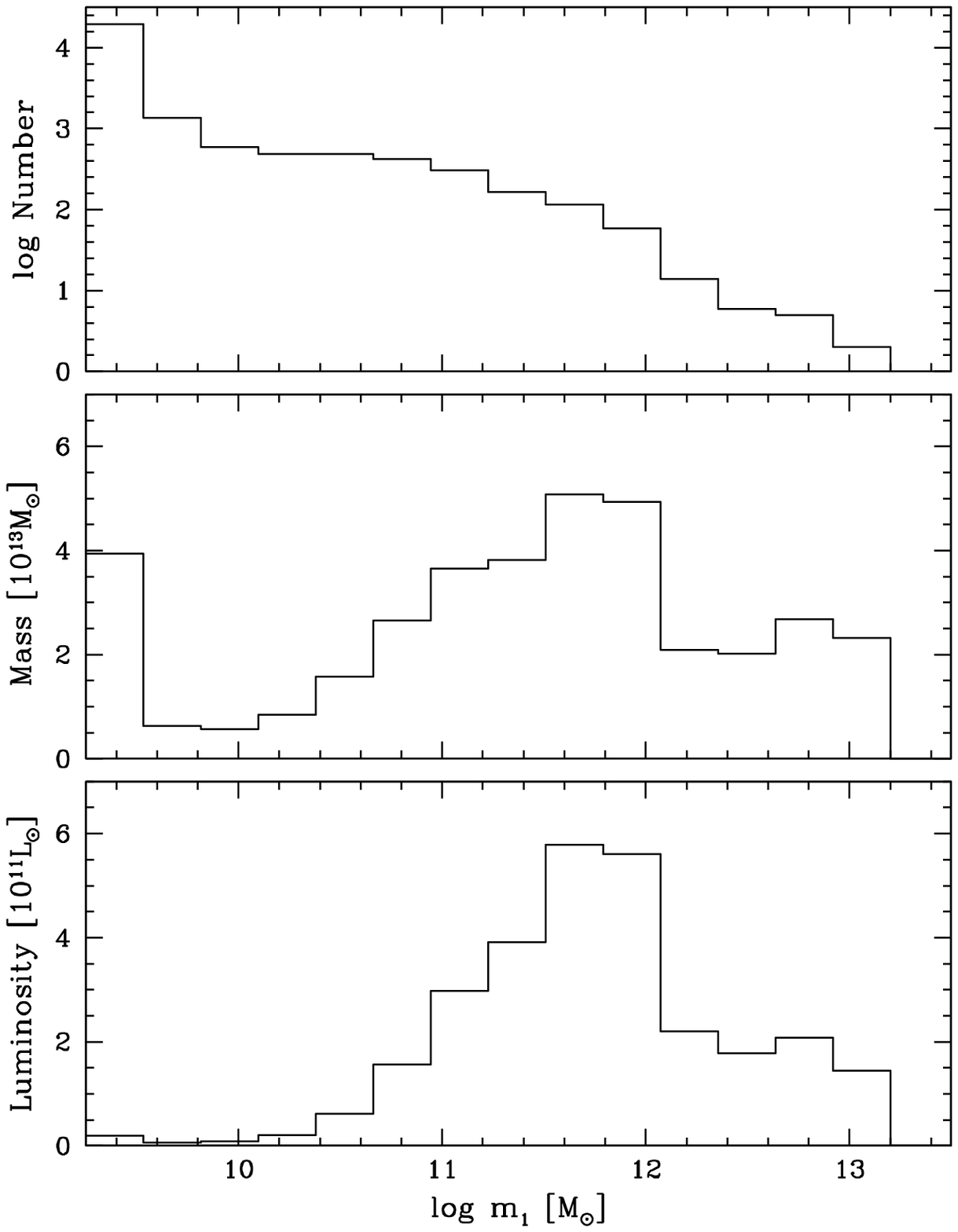}
\caption{
Properties of tidally destructed galaxies which contribute to the ICL,
versus mass. Top panel:
number of galaxies in each mass bin; middle panel: total mass in each mass bin;
bottom panel: total luminosity in each mass bin.}
\label{destruct_histo}
\end{center}
\end{figure}

\begin{figure}
\begin{center}
\includegraphics[width=6.5in]{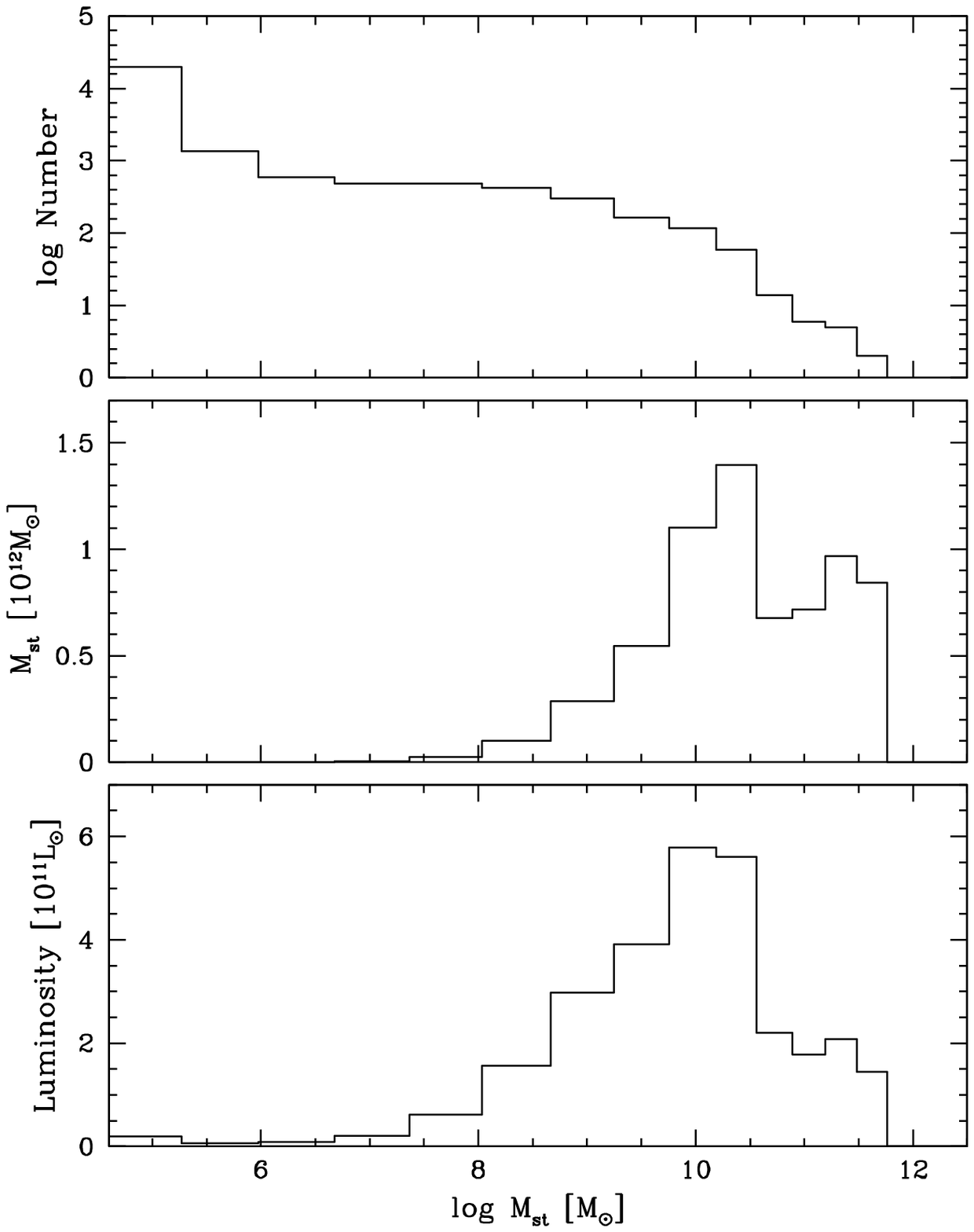}
\caption{
Properties of tidally destructed galaxies which contribute to the ICL,
versus stellar mass. Top panel:
number of galaxies in each mass bin; middle panel: total stellar
mass in each mass bin;
bottom panel: total luminosity in each mass bin.}
\label{destruct_histo_ms2}
\end{center}
\end{figure}

Figures~\ref{destruct_histo} and~\ref{destruct_histo_ms2}
show the properties of the galaxies that are
tidally destroyed, with fragments dispersed into the intracluster space,
as functions of the galaxies' total mass~$m_1$ (Fig.~\ref{destruct_histo})
and stellar mass~$M_{\rm st}$ (Fig.~\ref{destruct_histo_ms2}).
The top panels shows the number of galaxies destroyed in
different mass bins. Most galaxies destroyed are low-mass galaxies: 89.8\% 
of galaxies destroyed have masses $m_1<10^{10}\msun$,
$M_{\rm st}<2.8\times10^6\msun$, while only 0.16\%
have masses $m_1>10^{12}\msun$, $M_{\rm st}>3\times10^{10}\msun$.
The middle panels shows the total mass and
total stellar mass in
each mass bin. The contribution of low-mass galaxies is very small. 
42.1\% of 
the mass in tidal fragments comes from galaxies in the range 
$m_1=10^{11}-10^{12}\msun$ and $M_{\rm st}=6\times10^8-3\times10^{10}\msun$.
The peak at $m_1=M_{\min}=2\times10^9\msun$ results from 
the fact that there are too many galaxies of that mass to start with,
as we explained in \S~3.1. Using equation~(\ref{ml1}), we calculated the total
luminosity in each mass bin. The resulting distribution is
plotted in the bottom panels of Figures~\ref{destruct_histo}
and~\ref{destruct_histo_ms2}.
57.9\% of the ICL comes from galaxies in the range $m_1=10^{11}-10^{12}\msun$,
$M_{\rm st}=6\times10^8-3\times10^{10}\msun$,
30.6\% from massive, $m_1>10^{12}\msun$, $M_{\rm st}>3\times10^{10}\msun$ 
galaxies, and only 11.5\%
from low-mass, $m_1<10^{11}\msun$, $M_{\rm st}<6\times10^8\msun$ galaxies, 
including 1.0\% from
$m_1<10^{10}\msun$, $M_{\rm st}<2.8\times10^6\msun$ galaxies. 
\citet{willmanetal04} also found that 
intermediate-mass galaxies were an important contributor to the ICL.

\subsection{Cluster Analysis}

\subsubsection{Intracluster Light Fraction}

We identify clusters using a standard friends-of-friends algorithm (FOF)
with a linking length equal to 0.25 times the mean particle spacing
(corresponding to $48.8\,{\rm kpc}$ comoving). 
This algorithm identifies the dark matter particles, galaxies, and
tidal fragments that belong to each cluster. The term {\it tidal fragment\/}
refers here to galaxies that have been flagged as being tidally destroyed,
with their fragments dispersed into the intracluster space.
At $z=0$, we find
18 massive clusters, with masses $M_{\rm cl}>10^{14}\msun$. 
For each galaxy and tidal fragment located in these clusters, 
we calculate the luminosity using 
equation~(\ref{ml1}). By adding these luminosities, we get the total 
galaxy luminosity $L_{\rm gal}$ and the total intracluster luminosity
$L_{\rm ICL}$ for each cluster.
The properties of the clusters are listed in Table~\ref{table_cl}.
$M_{\rm cl}$, $M_{\rm gal}$, and $M_{\rm tid}$ are the total mass
of the cluster, the mass in galaxies, and the mass in tidal fragments,
respectively. $L_{\rm gal}$ and $L_{\rm ICL}$ are the galaxy and intracluster
luminosity, respectively, and 
$f_{\rm ICL}\equiv L_{\rm ICL}/(L_{\rm gal}+L_{\rm ICL})$ is the fraction of
intracluster light.

The values of $f_{\rm ICL}$ vary from 1\% to 58\%, while
observed values vary from 1.6\% to 50\% (see Table~12 of Paper~I). 
Our simulations therefore reproduce
the range of observed values of $f_{\rm ICL}$. However, we only have 
4 clusters (out of 18) with $f_{\rm ICL}<20\%$, while such low values are more
common among observed clusters. \citet{rmmb11} simulated 6 clusters, and
found values of $f_{\rm ICL}$ in the range $9\%-36\%$.
Most of their simulated
clusters have $f_{\rm ICL}<20\%$\footnote{Each cluster in the simulations
of \citet{rmmb11} has several values of $f_{\rm ICL}$ because they experiment
with various techniques for calculating that quantity.}. \citet{hbt08}
report median values of $f_{\rm ICL}=20\%$ for halos with masses 
$M_{\rm cl}\sim10^{13}h^{-1}\msun$ and $f_{\rm ICL}=30\%$ for halos with masses 
$M_{\rm cl}\sim10^{15}h^{-1}\msun$. Visual inspection of their Figure~6 indicates 
values of $f_{\rm ICL}$ in the range $10\%-50\%$ for the mass range
$M_{\rm cl}>10^{14}\msun$ we consider.
Overall, there is a
broad agreement between the values of $f_{\rm ICL}$ obtained by us, by
\citet{rmmb11}, by \citet{hbt08}, and the observed values.
The ranges of values are very wide. In \S~3.3.3 below, we investigate
the physical origin of this, and argue that it is a consequence of the 
hierarchical formation of clusters.

\begin{deluxetable}{lcccccc}
\tablecolumns{7}
\tablecaption{Properties of Massive Clusters}
\tablewidth{0pt}
\tablehead{
\colhead{Name} & 
\colhead{$M_{\rm cl}$ $[10^{14}\msun]$} &
\colhead{$M_{\rm gal}$ $[10^{14}\msun]$} & 
\colhead{$M_{\rm tid}$ $[10^{14}\msun]$} & 
\colhead{$L_{\rm gal}$ $[10^{11}\lsun]$} & 
\colhead{$L_{\rm ICL}$ $[10^{11}\lsun]$} & 
\colhead{$f_{\rm ICL}$ $[\%]$}
}
\startdata
C01 & 12.16 & 0.869 & 0.539 & 3.887 & 4.207 & 52 \cr
C02 & 10.28 & 0.914 & 0.313 & 3.289 & 2.151 & 40 \cr
C03 &  5.62 & 0.573 & 0.110 & 2.804 & 0.879 & 24 \cr
C04 &  5.40 & 0.369 & 0.223 & 2.156 & 1.866 & 46 \cr
C05 &  3.34 & 0.228 & 0.169 & 1.071 & 1.504 & 58 \cr
C06 &  2.98 & 0.315 & 0.037 & 1.916 & 0.256 & 12 \cr
C07 &  2.52 & 0.204 & 0.096 & 1.279 & 0.813 & 39 \cr
C08 &  2.02 & 0.192 & 0.045 & 1.204 & 0.401 & 25 \cr
C09 &  1.94 & 0.187 & 0.041 & 1.236 & 0.317 & 20 \cr
C10 &  1.71 & 0.130 & 0.072 & 1.019 & 0.668 & 40 \cr
C11 &  1.63 & 0.146 & 0.040 & 1.077 & 0.351 & 25 \cr
C12 &  1.36 & 0.094 & 0.066 & 0.719 & 0.644 & 47 \cr
C13 &  1.23 & 0.124 & 0.021 & 0.795 & 0.162 & 17 \cr
C14 &  1.21 & 0.105 & 0.032 & 0.828 & 0.277 & 25 \cr
C15 &  1.13 & 0.103 & 0.033 & 0.669 & 0.312 & 32 \cr
C16 &  1.11 & 0.117 & 0.003 & 0.745 & 0.006 &  1 \cr
C17 &  1.09 & 0.108 & 0.012 & 0.702 & 0.087 & 11 \cr
C18 &  1.06 & 0.162 & 0.053 & 1.234 & 0.497 & 29 \cr
\enddata
\label{table_cl}
\end{deluxetable}

\begin{figure}
\begin{center}
\includegraphics[width=7in]{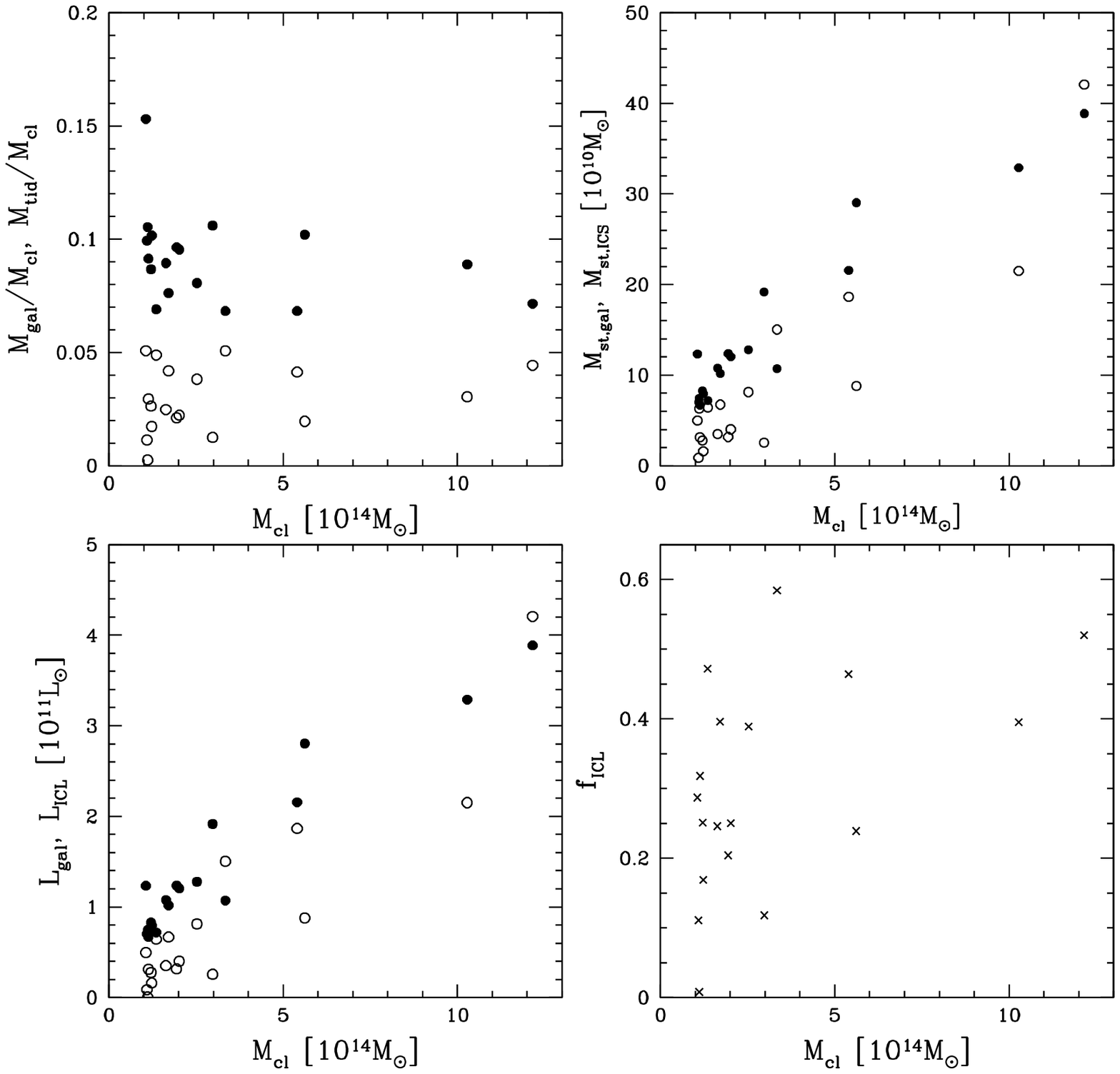}
\caption{
Top left panel: fraction of cluster mass inside galaxies
(solid circles) and inside tidal fragments (open circles)
versus cluster mass.
Top right panel: galaxy stellar mass
(solid circles) and intracluster stellar mass (open circles)
versus cluster mass.
Bottom left panel: galaxy luminosity
(solid circles) and intracluster luminosity (open circles)
versus cluster mass.
Bottom right panel: intracluster light fraction versus cluster mass.}
\label{clusters}
\end{center}
\end{figure}

Figure~\ref{clusters} shows the dependencies of cluster properties 
on the total mass of the cluster. The top left panel shows the mass fraction
in galaxies and tidal fragments. There is a lot of scatter, but overall
the mass fraction is around 0.08 for galaxies and 0.03 for tidal fragments,
with $M_{\rm gal}>M_{\rm tid}$ for all clusters. The top right panel shows
the stellar masses
$M_{\rm st,gal}$ and $M_{\rm st,ICS}$ in galaxies and intracluster stars, 
and the bottom left panel shows
the luminosities
$L_{\rm gal}$ and $L_{\rm ICL}$. All these quantities 
increase roughly linearly with $M_{\rm cl}$.
Most of the light comes from the galaxies, with two notable exceptions: 
clusters C01 (the most massive one) and C05. The bottom right panel shows the
intracluster light fraction $f_{\rm ICL}$. There is no obvious correlation
with cluster mass, except for the fact that massive clusters tend to have
large values of $f_{\rm ICL}$, with 4 of the 5 most massive clusters having
$f_{\rm ICL}\leq40\%$, while only 2 of the least 13 massive ones have
values of $f_{\rm ICL}$ this high. 
Some studies have found no significant dependence of $f_{\rm ICL}$ on cluster
mass \citep{dmb10,pssd10,rmmb11}, while others found that $f_{\rm ICL}$ tends
to increase with $M_{\rm cl}$ \citep{pbz07,muranteetal07,hbt08}. Our results are
somehow intermediate. We do not find very massive clusters with low 
$f_{\rm ICL}$, but we do find some less-massive clusters with high
$f_{\rm ICL}$.

\subsubsection{Cluster Evolution}

\begin{figure}
\begin{center}
\includegraphics[width=7in]{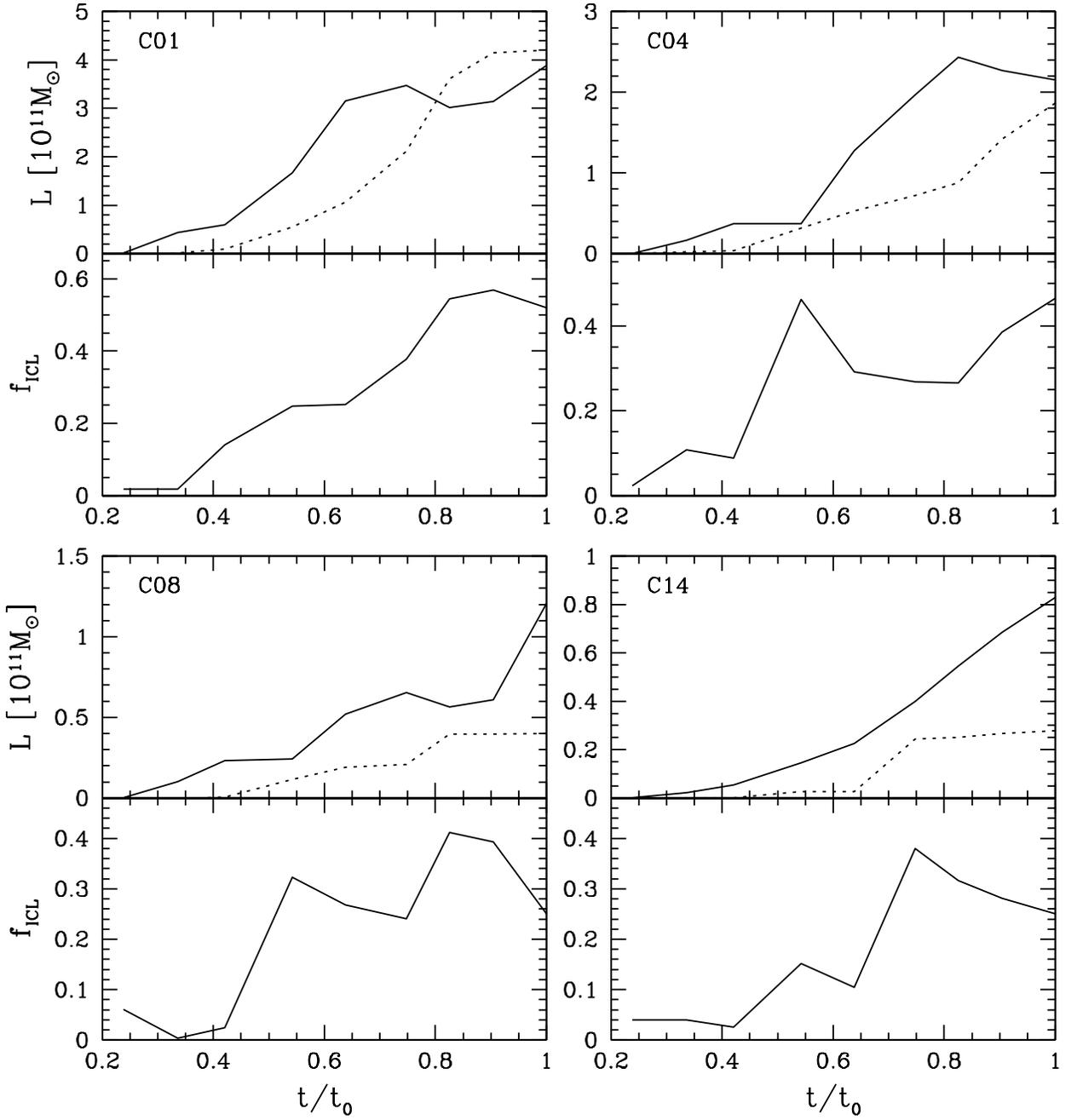}
\caption{
Time-evolution of the galaxy luminosity (top panels, solid curves), 
intracluster luminosity
(top panels, dotted curves), and intracluster light fraction $f_{\rm ICL}$
(bottom panels), for
clusters C01, C04, C08, and C14. $t_0$ is the present age of the universe.}
\label{evol4}
\end{center}
\end{figure}

\begin{figure}
\begin{center}
\includegraphics[width=7in]{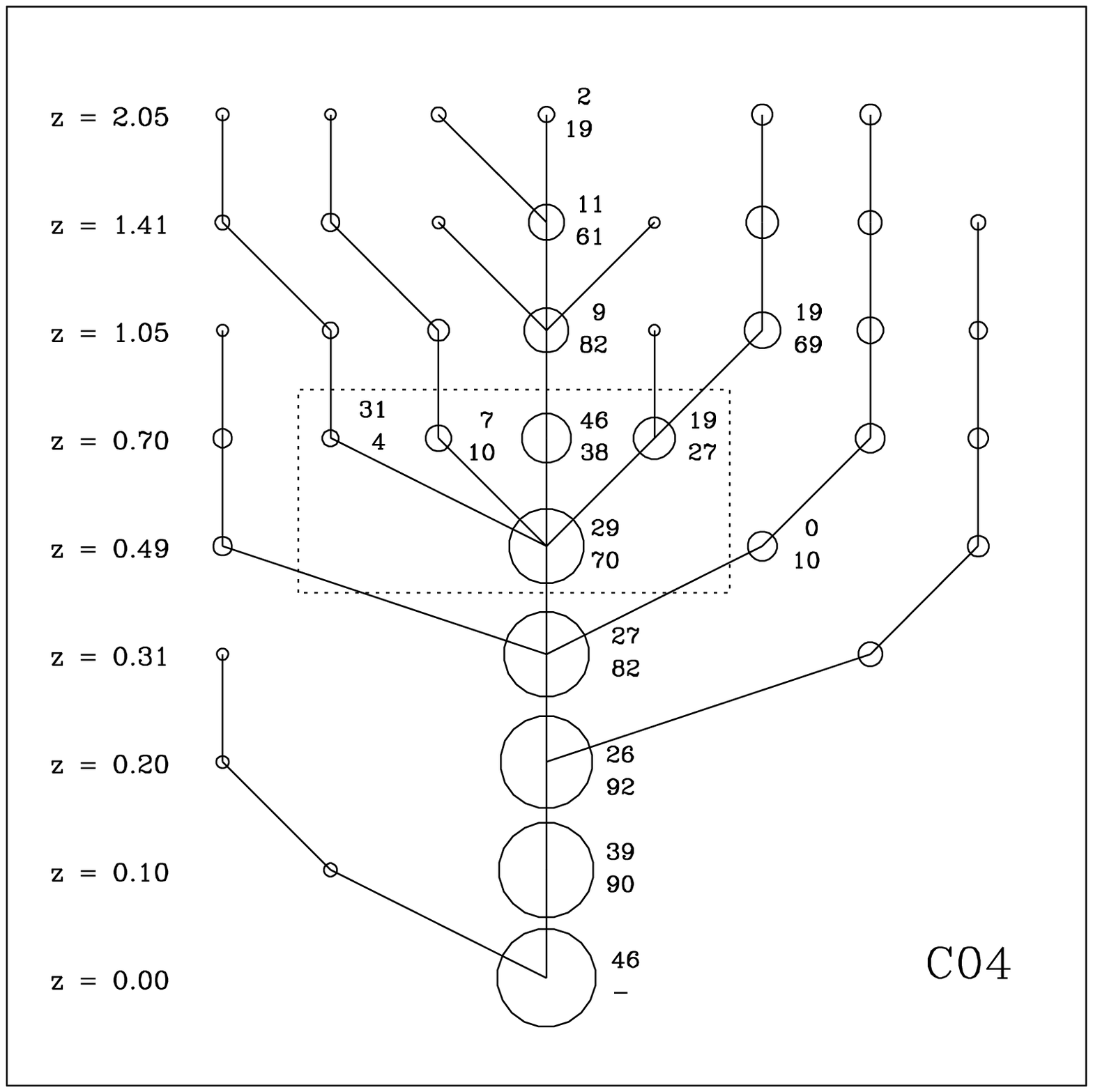}
\caption{
Merger tree for cluster C04. The area of each circle is proportional to the 
mass of the cluster. The numbers next to some clusters indicate the value
of $f_{\rm ICL}$ in percentage (top), and the relative 
contribution of that cluster to the mass 
of the merger remnant, also in percentage (bottom). 
The dotted box indicates a major merger, 
where the most massive progenitor contributes only 38\% of the final mass of 
the merger remnant. Redshifts are shown on the left.
For clarity, only clusters with $20,000$ particles or
more are shown.}
\label{mergertree}
\end{center}
\end{figure}

We used our FOF algorithm to build clusters catalogs at various
redshifts, and combined these catalogs to build merger trees
for all 18 massive clusters found at $z=0$. We then traced the ancestry
of each cluster back in time, following the most massive progenitor.
Figure~\ref{evol4} shows the evolution of the galaxy luminosity $L_{\rm gal}$,
intracluster luminosity $L_{\rm ICL}$, and intracluster light fraction 
$f_{\rm ICL}$, for a subset of 4 clusters. $L_{\rm ICL}$ increases with time
as galaxies get tidally destructed. It could only decrease if tidal fragments
were ejected from the clusters, but this does not seem to ever happen.
$L_{\rm gal}$ is affected by several processes. Galaxy formation increases
$L_{\rm gal}$, while tidal destruction followed by dispersal decreases it.
Galaxy merger events, and tidal destruction followed by accretion, both replace
two galaxies by one with the  same total mass. Since $L$ does not vary
linearly with $M$, the new galaxy does not have the total luminosity
of its two progenitors, as we explained in \S3.1. Overall, the 
values of $f_{\rm ICL}$ tend to increase with time,
in agreement with the results of \cite{rmmb11}. However, there can be sudden
drops in the value of $f_{\rm ICL}$, such as the one seen at $t/t_0=0.6$
in cluster C04. These drops are caused by major cluster mergers. When
two clusters of comparable masses, but with very different values of
$f_{\rm ICL}$, merge, the merger remnant will have a value of $f_{\rm ICL}$
that is intermediate between the values for the progenitors. If the 
progenitor
with the largest value of $f_{\rm ICL}$ was identified as the main progenitor,
then the net effect of the merger is to decrease $f_{\rm ICL}$ for that cluster.
To illustrate this, we show in Figure~\ref{mergertree} the merger tree for 
cluster
C04. Though the cluster was built mostly through a series a minor merger
(where one progenitor provides 80\% or more of the mass), we see that a major
merger took place between redshifts $z=0.70$ and $0.49$. The main progenitor has
a value $f_{\rm ICL}=0.46$, but provides only 38\% of the mass. The next three
progenitors all have lower values of $f_{\rm ICL}$, and together provide
41\% of the mass. As a result, the value of $f_{\rm ICL}$ drops from 0.46
to 0.29. A similar thing happens at $t/t_0=0.95$ for cluster C08.
In the cases of clusters C01 and C14, we found that the
late drop in $f_{\rm ICL}$ was not caused by a major merger, but rather by
a sudden increase in galaxy formation.

These results are consistent with the results of \citet{rmmb11}. Their 
simulations show that $f_{\rm ICL}$ does not increase at a constant rate, 
and does not always vary monotonically.

\subsubsection{The Range of Intracluster Light Fractions}

\begin{deluxetable}{lcrccccccc}
\tablecolumns{10}
\tablecaption{Main Ancestry of Massive Clusters}
\tablewidth{0pt}
\tablehead{
\colhead{Name} & \colhead{$f_{\rm ICL}^{\phantom1}\ [\%]$} &
\colhead{$z=0.10$} & \colhead{0.20} & \colhead{0.31} & 
\colhead{0.49} & \colhead{0.70} & \colhead{1.05} & \colhead{1.41} & 
\colhead{2.06}
}
\startdata
C01 & 52 & 90 & 90 & 86 & 80 & 57 & {\bf[35]} & 79 & {\bf[22]} \cr
C02 & 40 & 89 & 92 & 87 & 78 & 84 & {\bf[35]} & 80 & {\bf[31]} \cr
C04 & 46 & 90 & 92 & 82 & 70 & {\bf[38]} & 82 & 61 & {\bf[19]} \cr
C05 & 58 & 92 & 93 & 90 & 81 & 87 & {\bf[46]} & {\bf[32]} & {\bf[29]} \cr
C07 & 39 & 79 & 91 & 88 & 75 & 83 & 70  & {\bf[33]} & 69 \cr
C08 & 25 & 64 & 90 & 92 & 87 & 57 & 80 & {\bf[45]} & 71 \cr
C10 & 40 & 91 & 91 & 73 & {\bf[48]} & 53 & 67 & 66 & {\bf[43]} \cr
C12 & 47 & 90 & 92 & 83 & {\bf[45]} & {\bf[48]} & 67 & 63 & \nodata \cr
C14 & 25 & 90 & 87 & 85 & {\bf[41]} & 75 & {\bf[47]} & 51 & 71 \cr
\hline
C03 & 24 & 88 & 90 & 88 & 63 & 80 & 53 & 75 & {\bf[18]} \cr 
C06 & 12 & 89 & 90 & 70 & 81 & 84 & 64 & 77 & 55 \cr
C09 & 20 & 89 & 52 & 90 & 80 & 85 & 63 & 73 & {\bf[25]} \cr
C11 & 25 & 89 & 87 & 56 & 84 & 83 & 70 & 73 & {\bf[26]} \cr
C13 & 17 & 92 & 73 & 85 & 81 & 76 & 75 & 73 & 57 \cr
C15 & 32 & 92 & 87 & 91 & 84 & 69 & 77 & 69 & {\bf[27]} \cr
C16 &  1 & 89 & 91 & 86 & 85 & 82 & 71 & 76 & 74 \cr
C17 & 11 & 86 & 90 & 89 & 81 & 84 & 69 & 74 & {\bf[30]} \cr
C18 & 29 & 87 & 69 & 91 & 88 & 68 & 83 & 52 & 54 \cr
\enddata
\label{table_ancestry}
\end{deluxetable}

Once the merger trees of clusters are built, we can investigate the origin
of the very wide range in the values of $f_{\rm ICL}^{\phantom1}$ (from 1\% to
58\% in our simulation, and a comparable range in the observed values).
For each cluster, we started at redshift $z=0$, and followed the ancestry
back in time along the main progenitors. The results are shown in 
Table~\ref{table_ancestry}. For each cluster, we indicate, at each redshift 
$z>0$ the contribution in percentage of that cluster to the mass of the merger
remnant at the next redshift (the reader will recognize, for cluster C04,
the numbers plotted along the central line in Fig.~\ref{mergertree}).
We indicated in boldface and square brackets the major mergers, when 
less than 50\% of the mass of the merger remnant comes from the
main progenitor. We also separated the clusters in two groups (top and bottom).
The clusters in the top group all experienced a major merger at intermediate
redshift $z<1.41$. The clusters in the bottom group experienced no major merger,
or a major merger at high redshift. There is a striking correlation between the
presence of major mergers at intermediate redshifts and the final value
of $f_{\rm ICL}^{\phantom1}$. All clusters in the top group have 
$f_{\rm ICL}^{\phantom1}\geq25\%$, and the seven highest values of 
$f_{\rm ICL}^{\phantom1}$
are found in that group; all clusters in the bottom group have 
$f_{\rm ICL}^{\phantom1}\leq32\%$, and the six lowest values of
$f_{\rm ICL}^{\phantom1}$
are found in that group. Clearly, major mergers at intermediate redshift
drive the evolution of $f_{\rm ICL}^{\phantom1}$ and determine the final value
of $z=0$. While minor mergers will tend to leave clusters essentially
undisturbed, major mergers can have dramatic effects. In particular, the
merger of two clusters of comparable masses, which contain comparable
numbers of galaxies, can lead to a sudden increase in the number of
density of galaxies. Since the frequency of encounters scales like the square
of that number density, we expect a significant increase in the
rate of encounters immediately after
a major merger, and that rate might remain high all the way to 
$z=0$. This will result in a large number of tidal destruction events, and a
correspondingly high value of $f_{\rm ICL}^{\phantom1}$. Notice that major mergers
of clusters at $z=2.06$ and do not have much effect, because
the clusters do not contain many galaxies at that time.

This result is not at odds with the conclusion of the previous
section. A major merger between clusters of comparable masses
and different values of $f_{\rm ICL}^{\phantom1}$ 
can lead to a sudden drop in the value of $f_{\rm ICL}^{\phantom1}$, 
depending on which progenitor is identified as
the main one. But this sudden drop can be more than compensated by
the increase rate of encounters that take place after the merger
and can last all the way to the present. Cluster C04 provides a
good illustration of this. $f_{\rm ICL}^{\phantom1}$ drops from 46\% 
to 29\% during the major merger, but is back at 46\% by $z=0$. 

\subsubsection{Projected Luminosity Distribution}

\begin{figure}
\begin{center}
\includegraphics[width=6in]{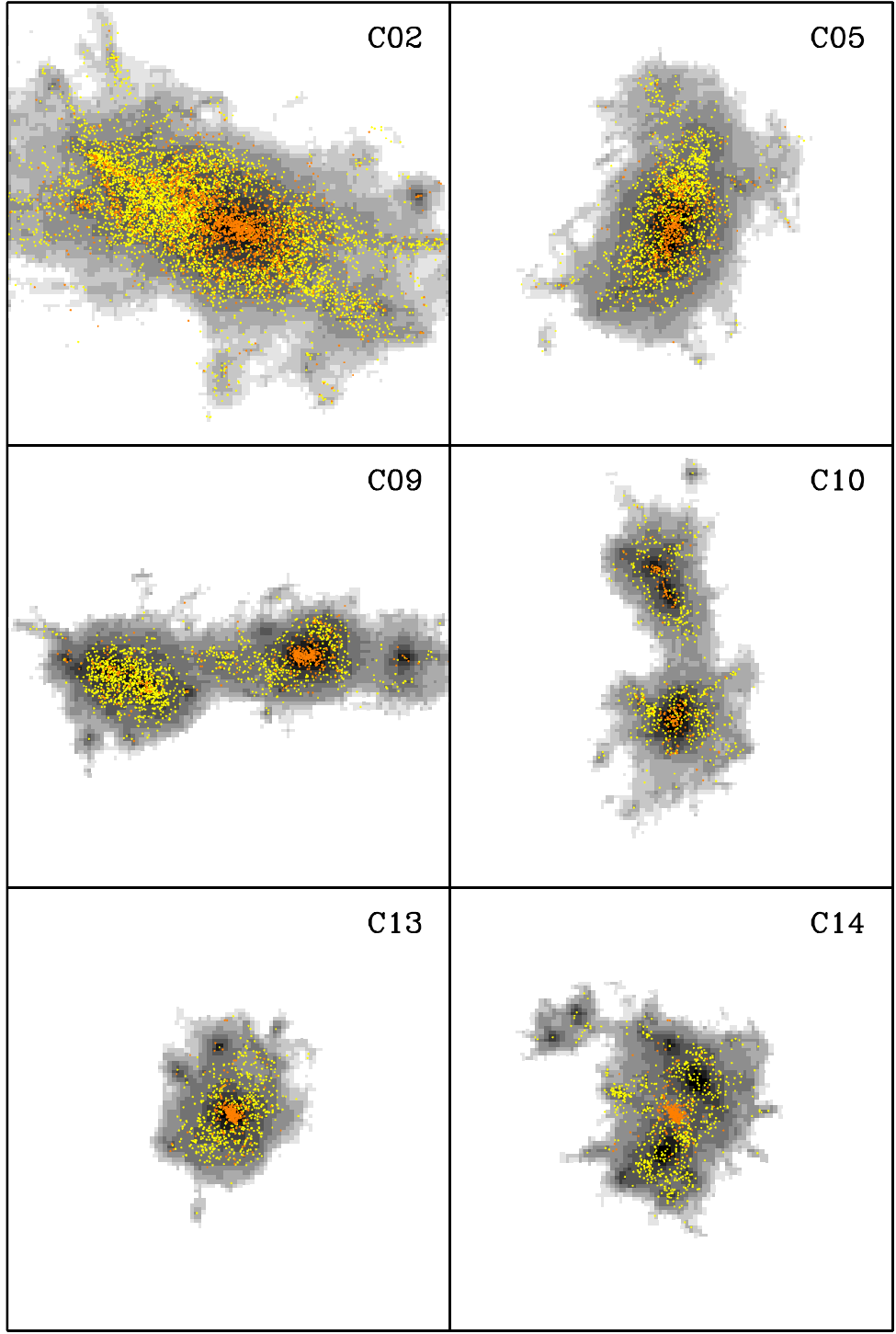}
\caption{Projected maps of some clusters. The greyscale shows the projected
surface density of dark matter (shades are separated by 0.3~dex).
Yellow and orange dots show galaxies and tidal fragments, respectively.
All panels are $8\,{\rm Mpc}\times8\,{\rm Mpc}$.}
\label{maps}
\end{center}
\end{figure}

\begin{figure}
\begin{center}
\includegraphics[width=6.5in]{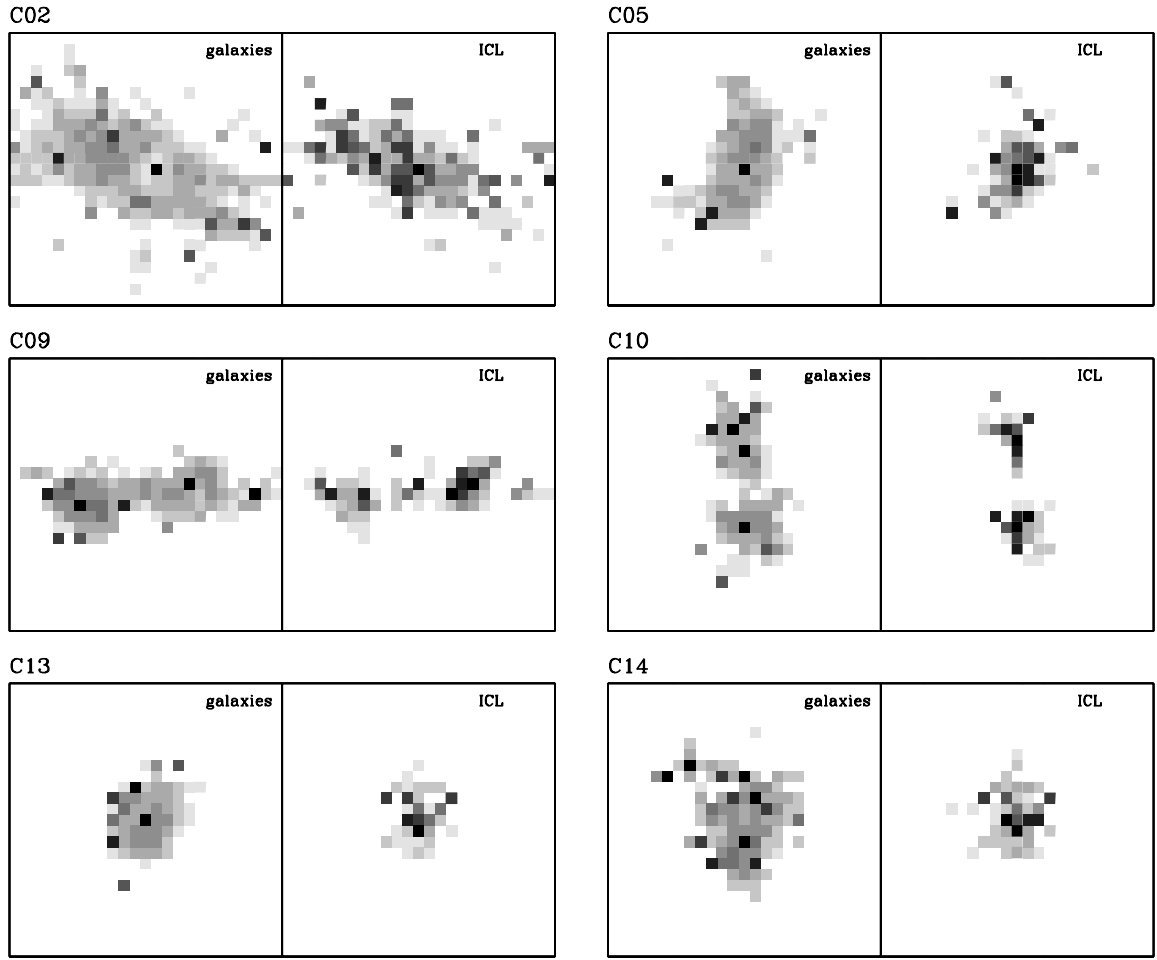}
\caption{Projected luminosity maps of some 
clusters. The greyscale shows the projected
luminosity density of galaxies (left panel for each cluster) and
intracluster light (right panel for each cluster).
Shades are separated by 0.5~dex, and for each
cluster, galaxies luminosity and ICL are plotted using the same greyscale.
All panels are $8\,{\rm Mpc}\times8\,{\rm Mpc}$.}
\label{maps_lum}
\end{center}
\end{figure}

\begin{figure}
\begin{center}
\includegraphics[width=6.5in]{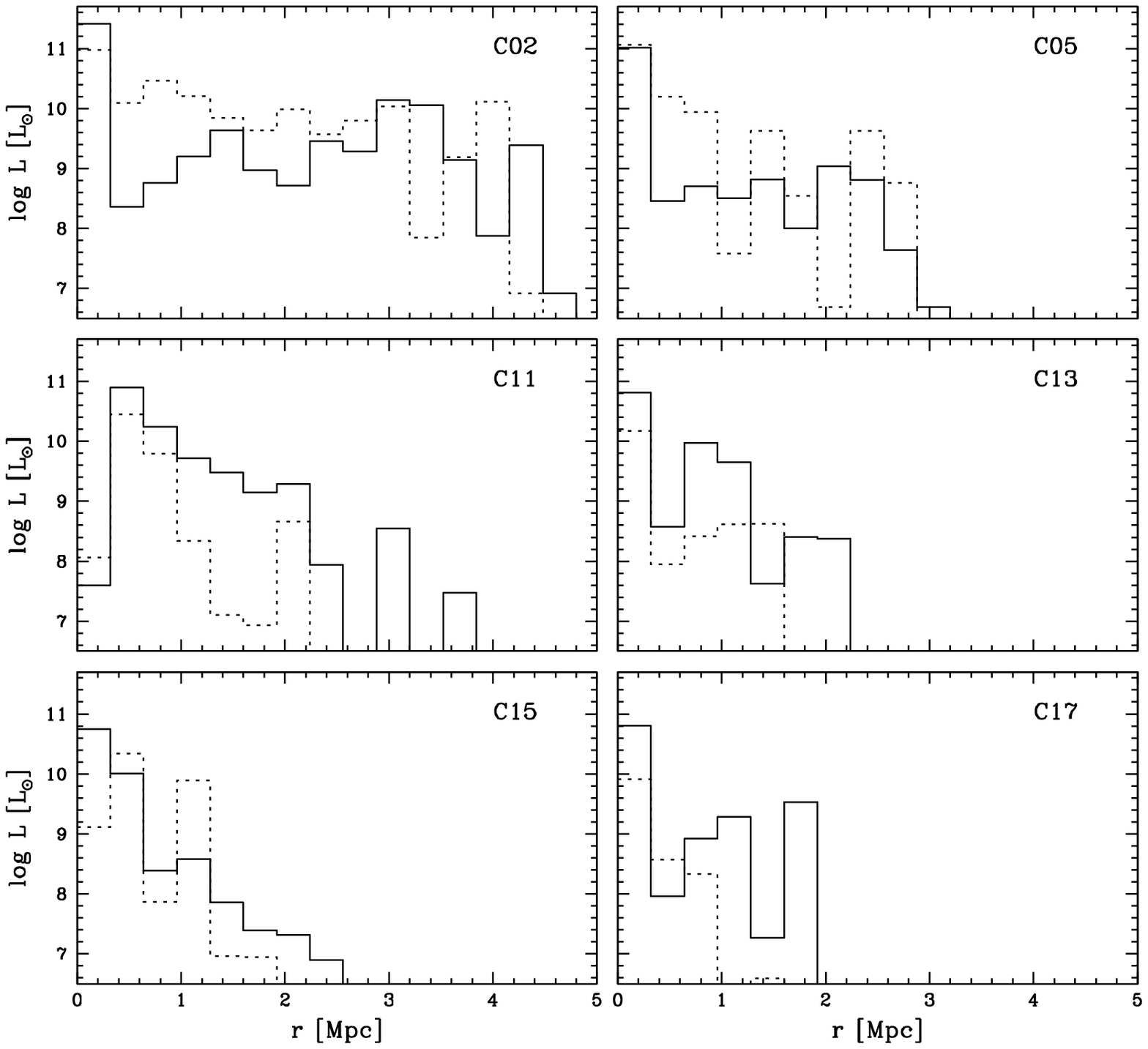}
\caption{Luminosity in projected radial bins for some clusters.
Solid and dotted lines show the galactic luminosity
and intracluster luminosity, respectively.}
\label{maps_reduce}
\end{center}
\end{figure}

Figure~\ref{maps} 
shows the structure of some of the clusters. The greyscale images
show the projected surface density of dark matter. The yellow and orange dots
show the galaxies and tidal fragments, respectively. Some clusters, like C05
and C13, are relaxed objects with a well-defined core where the density 
of dark matter and the number density of galaxies and tidal fragments all
peak. Clusters like C09 and C10 are clearly two clusters undergoing mergers,
and each one has two separate cores. Interestingly, in the case of C09,
one core has a large number of tidal fragments, while the other core has
very few. In all cases, we notice that the tidal fragments are much more
centrally concentrated than the galaxies. We calculated separately the
galaxy and interstellar luminosities for all clusters. The results are shown
in Figure~\ref{maps_lum}, 
for the same subset of clusters as in Figure~\ref{maps}. 
It is clear that, within each
core, the intracluster light is more concentrated than the galactic light.

We selected all clusters that have a single dominant core.
For each cluster, we 
calculated the position of the projected center-of-light,
both for the galactic and ICL components. We then
calculated the luminosity of each component in projected radial
bins of width $320\,{\rm kpc}$. We show some of the results in
Figure~\ref{maps_reduce}. The central galaxy luminosity tends to be dominated
by a bright central galaxy\footnote{As we explained in \S~3.1, that ``galaxy''
is actually several galaxies represented by one particle.}.
Cluster C11 has two bright galaxies located on opposite sides of
the center, which explains the low luminosity of the central bin.
The profiles show large variations due to the clumpiness of the
projected luminosity distributions, but there are some clear trends.
For the massive clusters C02 and C05, if we exclude the central bin,
the histograms tend to be flat for the galactic luminosity and decreasing 
for the intracluster luminosity, indicating that the latter is more
centrally concentrated. This is even clearer for the lower-mass clusters 
C11, C13, C15, and C17. Both the galactic luminosity and intracluster 
luminosity decrease with radius, but the length scale of the intracluster
light is of order $0.5-0.7$ of the length scale of the galactic light. 
As we saw in Figure~\ref{destruct_histo}, 
the destruction of intermediate-mass galaxies
is the primary contributor to the intracluster light. These galaxies can
only be destroyed by more massive ones, and these
these massive galaxies tend to reside in the center of clusters.

Notice that, in our algorithm, tidal fragments are treated as single
particles. When a galaxy is tidally destroyed, the algorithm flags it as
a tidal fragment with the same mass. Intermediate-mass tidal fragments
might be subject to mass segregation, which in this case would be a spurious
algorithmic effect, leading to an overestimate of the concentration of ICL
in the central region. Still, if most tidal fragments are produced
by interactions taking place in the center of clusters, we do expect the 
fragments to be found in these regions at later time, whether they are
represented by one massive particles or several less massive ones.

\section{DISCUSSION}

Our model combines a gravitational N-body algorithm, a subgrid treatment
of galaxy formation, galaxy mergers, and tidal destruction, and a fitting
formula for $M/L$ vs. $M$ derived from observations.
The greatest virtue of this model is its simplicity. By only including 
gravitational dynamics, and by using a subgrid treatment at small mass scales,
we are able to perform a large cosmological simulation at relatively low 
computational cost.
We simulate a cubic volume of $100\,{\rm Mpc}$ in size, containing 18
clusters of mass above $10^{14}\msun$. Doing a full hydrodynamical simulation
of this size, without subgrid physics, that could resolve in detail
galactic mergers and tidal destruction at the dwarf galaxy scale
would be prohibitive. Our simulation only took a few weeks on a 
16-processor computer.

There are some caveats implied by the method we use. By only including
gravity, and not hydrodynamics, our algorithm can describe the structure and
evolution of the mass distribution in the universe, but not how this mass
is converted to light. Hence, the algorithm cannot predict the luminosity
of galaxies self-consistently. 
This is why we used the observed relation between $M/L$
and $M$ [eq.~(\ref{ml1})] to calculate the luminosity of galaxies and 
tidal fragments. Actually, this relation gives the {\it average} value of
$M/L$. For any particular value of $M$, there is a distribution of values of
$M/L$ \citep{ymvdb03}. 
We ignored this, and calculated the luminosity $L$ of an object
of mass $M$ by using equation~(\ref{ml1}) directly. We could instead have
used the actual distribution of values of $M/L$ and draw a random value from
that distribution for each object, but that would have been an overkill.
We formed $1,088,797$ dwarf galaxies in our simulation, and by $z=0$,
each massive cluster ($M>10^{14}\msun$) contains between 311 and $7,765$ 
galaxies, and between 91 and $2,113$ 
tidal fragments. Drawing the ratios $M/L$ from
a distribution instead of using the average value would hardly make any
difference in the total luminosity of these components, and the inferred
intracluster light fraction $f_{\rm ICL}$.

Using a subgrid model for galaxy mergers and tidal destruction enables us
to reach dwarf-galaxy scales while simulating a large cosmological volume,
but there is also another advantage: it provides an unambiguous determination
of the outcome of each close encounter (merger, tidal destruction, or
simple encounter). If the encounters were actually simulated in details, the 
determination of their outcome would require detailed analysis,
as some of the matter would merge, while some would be ejected, and
some of that ejected matter would be reaccreted. As we
argue here, and also in Paper~I, our subgrid model would not be
appropriate to describe a single encounter, but can correctly describe,
in a statistical way, the collective effect of a large number of
encounters. In the simulation presented in this paper, 
$590,262$ encounters resulted in
a merger, $8,314$ encounters resulted in tidal destruction followed
be dispersion, and $113,132$ encounters
resulted in tidal destruction followed by accretion. 

We have assumed that the intracluster light is caused entirely by
galaxies that are tidally destroyed. Some simulations suggest that
luminous matter can also be added to the intracluster space during mergers
(e.g. \citealt{muranteetal07}). Since we are neglecting this effect,
our values of $f_{\rm ICL}$ might be underestimated. However, we could argue
that a merger with some of the matter being dispersed in the ICL,
a case we do not consider, is an intermediate case between a tidal destruction
with complete dispersal of the fragments and a tidal destruction with
all the fragments being subsequently reaccreted, two cases we do consider.
Hence, two mergers with some of the matter dispersed in the intracluster
space could be
equivalent to a tidal destruction with complete dispersal of the
fragments plus a tidal destruction with complete reaccretion of the
the fragments.
Statistically, the net effect might be the same.

Our model has only one parameter that is truly
tunable: the coefficient $\Psi$ appearing in equation~(\ref{GF}). We adjusted
this value to reproduce the high-luminosity end of the luminosity function,
as seen in Figure~\ref{histo2}. Using a larger value of $\Psi$ might lead
to an improvement of the luminosity function at the low-luminosity 
end, but could worsen the fit at the high-luminosity end. Also, a larger
value of $\Psi$ would likely result in an increase in both $L_{\rm gal}$
and $L_{\rm ICL}$, while having a smaller effect on the value of $f_{\rm ICL}$.

\section{SUMMARY AND CONCLUSION}

We performed a numerical simulation of the formation of galaxies and
clusters, the destruction of galaxies by mergers and tides, and the
evolution of the galactic, extragalactic, and intracluster light,
inside a cosmological volume of size $(100\,\rm Mpc)^3$, in a $\Lambda$CDM
universe. Our main results are the following:

\begin{itemize}

\item Our simulation reproduces the observed
Schechter luminosity function for luminosities $L>10^{8.5}\lsun$, up to
$L=10^{11}\lsun$. We have a significant excess of galaxies at luminosities
$L<10^{6.5}\lsun$, and a deficit in the range $10^{6.5}\lsun-10^{8.5}\lsun$.
We attribute this to the discreteness of the galaxy masses. All galaxies
in our simulation have masses that are multiples of $M_{\min}=2\times10^9\msun$,
and this can affect the luminosity function at low $L$.

\item The number of mergers and tidal destruction events increase
exponentially with decreasing redshift, with mergers starting at $z\sim5.9$
and tidal destruction starting at $z\sim4.8$. This delay is caused by
the time it takes to build up galaxies of significantly different masses.
Mergers outnumber tidal destruction events by about an order of magnitude,
at all redshifts up to the present. When tidal destruction occurs,
dispersal of the fragments into the intracluster space
dominates over reaccretion of the
fragments by the larger galaxy, up to redshift $z\sim0.5$. This trend is then
reversed.

\item Tidal destruction is not limited to dwarf galaxies.
Intermediate-mass galaxies
and even high-mass galaxies are destroyed during encounters 
with even higher-mass galaxies. Tidal destruction and also mergers
involve galaxy pairs with all masses and all mass ratios. 
We found an interesting trend
for encounters between high-mass galaxies 
($M>3\times10^{11}\msun$, $M_{\rm st}>5\times10^9\msun$). 
The outcome of such encounter seems to depend almost entirely on
the mass ratio
between the galaxies. Small mass ratios ($m_2/m_1<1.2$) result in mergers,
intermediate mass ratios ($m_2/m_1<10$) result in tidal destruction with the
fragment being dispersed into the intracluster space, 
and higher mass ratios result in
tidal destruction, with the fragments being reaccreted by the massive galaxy.

\item Most galaxies destroyed by tides are low-mass galaxies. However,
the total luminosity provided by these low-mass galaxies is small.
57.9\% of the ICL comes from galaxies of
intermediate masses ($M=10^{11}\msun-10^{12}\msun$,
$M_{\rm st}=6\times10^8\msun-2\times10^{10}\msun$), while lower-mass galaxies
provide only 11.5\% of the ICL. Essentially, the bulk of
the ICL comes from galaxies of masses $m_1=10^{11}\msun-10^{12}\msun$ which are
tidally destroyed by galaxies of mass $m_2=(1.2-10.0)m_1$.
Higher mass ratios result in reaccretion of the tidal fragments.

\item The present intracluster light fraction $f_{\rm ICL}$ is in the range
$1\%-58\%$. This is consistent with observations, and with simulations 
presented by other groups. Even though mergers outnumber tidal destruction
events by an order of magnitude, 
the latter are sufficient to explain the observed ICL.
The galaxy luminosity $L_{\rm gal}$ and
intracluster luminosity $L_{\rm ICL}$ both increase with cluster mass.
The intracluster light fraction $f_{\rm ICL}$ does not show any particular
trend with cluster mass, except for the fact that we did not find massive
clusters with low values of $f_{\rm ICL}$.

\item The value of $f_{\rm ICL}$ for any particular cluster tends to increase
with time. However, some clusters experience sudden drops in $f_{\rm ICL}$, that
can happen at any redshift. At early times, these sudden drops are caused by
major mergers, when the cluster absorbs another cluster of comparable mass
but with a smaller value of $f_{\rm ICL}$. At late times, they are caused by 
a sudden increase in galaxy formation not accompanied by a corresponding 
increase in tidal destruction.

\item Several clusters are not relaxed at $z=0$, and show complex structures
with multiple cores. Focusing on the clusters with a well-defined core,
we found that the distribution of ICL is more concentrated
than the distribution of galactic light. Most of the ICL comes
from intermediate-mass galaxies destroyed by massive ones. Since 
these massive galaxies tend to reside in the center of clusters, this 
explains a relatively small extent of the ICL.

\end{itemize}

\acknowledgments

This work benefited from stimulating discussions with J. Navarro.
All calculations were performed at the Laboratoire
d'astrophysique num\'erique, Universit\'e Laval. 
We thank the Canada Research Chair program and NSERC for support.
PB acknowledges support from the FP7 ERC Starting Grant {\sl cosmoIGM\/}.
HM is thankful to the Department of Physics and Astronomy,
University of Victoria, for its hospitality. 

%

\end{document}